\documentclass[
  aps,
  prx,
  reprint,
  superscriptaddress,
  amsmath,amssymb,
  floatfix
]{revtex4-2}
\usepackage{CJK}
\usepackage[T1]{fontenc}
\usepackage[utf8]{inputenc} 

\usepackage{graphicx}
\usepackage{dcolumn}
\usepackage{bm}
\usepackage{multirow}
\usepackage{amsmath,amssymb,amsfonts}
\usepackage{amsthm}
\usepackage{mathrsfs}
\usepackage{physics}
\usepackage{xcolor}
\usepackage{textcomp}
\usepackage[normalem]{ulem}
\usepackage{booktabs}
\usepackage{algorithm}
\usepackage{algorithmicx}
\usepackage{algpseudocode}
\usepackage{listings}
\usepackage[per-mode=symbol]{siunitx}
\usepackage[colorlinks=true,citecolor=blue,linkcolor=blue,urlcolor=blue]{hyperref}
\usepackage{tikz}
\usetikzlibrary{arrows.meta,backgrounds,fit,calc}

\theoremstyle{plain}

\theoremstyle{definition}

\theoremstyle{remark}

\begin{document}
\title{Deep Reinforcement Learning for Individual Atomic Control and Cooling}

\author{Matthew L. Peters}
\thanks{These authors contributed equally.}
\affiliation{MIT-Harvard Center for Ultracold Atoms and Research Laboratory of Electronics, Massachusetts Institute of Technology, Cambridge, MA 02139, USA}
\affiliation{Department of Physics, Massachusetts Institute of Technology, Cambridge, MA 02139, USA}

\author{Guoqing Wang}
\thanks{These authors contributed equally.}
\affiliation{MIT-Harvard Center for Ultracold Atoms and Research Laboratory of Electronics, Massachusetts Institute of Technology, Cambridge, MA 02139, USA}
\affiliation{Department of Physics, Massachusetts Institute of Technology, Cambridge, MA 02139, USA}

\author{David C. Spierings}
\thanks{These authors contributed equally.}
\affiliation{MIT-Harvard Center for Ultracold Atoms and Research Laboratory of Electronics, Massachusetts Institute of Technology, Cambridge, MA 02139, USA}
\affiliation{Department of Physics, Massachusetts Institute of Technology, Cambridge, MA 02139, USA}

\author{Niv Drucker}
\affiliation{
   Quantum Machines, Tel Aviv-Yafo 6721407, Israel}

\author{Meng-Wei Chen}
\affiliation{MIT-Harvard Center for Ultracold Atoms and Research Laboratory of Electronics, Massachusetts Institute of Technology, Cambridge, MA 02139, USA}
\affiliation{Department of Physics, Massachusetts Institute of Technology, Cambridge, MA 02139, USA}

\author{Audrey Bartlett}
\affiliation{MIT-Harvard Center for Ultracold Atoms and Research Laboratory of Electronics, Massachusetts Institute of Technology, Cambridge, MA 02139, USA}
\affiliation{Department of Physics, Massachusetts Institute of Technology, Cambridge, MA 02139, USA}

\author{Isaac Chuang}
\email[]{ichuang@mit.edu}
\affiliation{MIT-Harvard Center for Ultracold Atoms and Research Laboratory of Electronics, Massachusetts Institute of Technology, Cambridge, MA 02139, USA}
\affiliation{Department of Physics, Massachusetts Institute of Technology, Cambridge, MA 02139, USA}
\affiliation{Department of Electrical Engineering and Computer Science, Massachusetts Institute of Technology, Cambridge, MA 02139, USA}

\author{Vladan Vuleti\'c}
\email[]{vuletic@mit.edu}
\affiliation{MIT-Harvard Center for Ultracold Atoms and Research Laboratory of Electronics, Massachusetts Institute of Technology, Cambridge, MA 02139, USA}
\affiliation{Department of Physics, Massachusetts Institute of Technology, Cambridge, MA 02139, USA}

\begin{abstract}

Real-time feedback control of quantum systems is often limited by partial observations, nonlinear dynamics and measurement noise, which make accurate model-based controllers difficult to design. Here we show that deep reinforcement learning can cool the motion of a single neutral atom coupled to a high-finesse optical cavity using only the continuously monitored cavity transmission. We first train the controller in simulation and then transfer it to the experiment, where online fine-tuning adapts it to unmodeled experimental dynamics. The learned policy damps the atom’s motion in real time and achieves a cooling time constant of \SI{388 \pm 14}{\micro s}, corresponding to only two motional periods in the trap. It also outperforms a standard linear differentiator controller in cooling speed while maintaining comparable atom retention over a broad range of operating conditions. These results establish reinforcement learning as a practical strategy for feedback control in quantum-limited experiments where compact analytical models are incomplete.

\end{abstract}

\maketitle

Individually trapped neutral atoms have emerged as a leading platform for quantum science, enabling breakthroughs in quantum simulation~\cite{FirstIsing_Browaeys_2016, FirstIsing_Lukin_2017}, metrology~\cite{MetrologicalAtom_Kaufman_2019, MetrologicalAtom_Endres_2025}, and the development of error-corrected quantum computers~\cite{QEC_AtomComputing_2024, QEC_Bluvstein_2023}. A powerful technique for enhancing control and measurement in these systems is to interface the atoms with a high-finesse optical cavity. Strong coupling between atom and cavity provides a channel for detecting an atom's internal state with high fidelity~\cite{CavityReadout_Deist_2022, CavityReadout_Grinkemeyer_2025}, mediating heralded entanglement~\cite{CavityReadout_Grinkemeyer_2025, CavityEntangle_Remple_2017, CavityEntangle_Dordevic_2021}, and serving as a sensitive probe for the atom's motional degrees of freedom~\cite{FallingAtom_Mabuchi_1996, FallingAtom_Hood_1998, FallingAtom_Munstermann_1999}. In particular, the ability to reconstruct classical atomic trajectories from transmitted cavity light~\cite{TrajectoryReconstruction_Rempe_2000, TrajectoryReconstruction_Mabuchi_1999, TrajectoryReconstruction_Hood_2000}
has enabled active control schemes such as feedback cooling, where precisely timed forces applied based on continuous observation systematically remove an atom’s kinetic energy ~\cite{FeedbackCooling_Rempe_2010}. This technique is broadly applicable to other systems, including nanoparticles~\cite{SubwavelengthCooling_Diehl_2018, ColdDamping_Tebbenjohanns_2019, DigitalParametricFeedback_Zheng_2019, FeedbackNanoparticle_Aikawa_2021}. 
Some examples use Kalman filtering to infer particle position, momentum, and drift from noisy measurement records, enabling trapping and cooling beyond immediate observation alone~\cite{Aspelmeyer_Kalman_2021, Kiesel_Kalman_2024}.
However, such model-based feedback schemes rely on an accurate, predefined physical model~\cite{FeedbackUntrapped_Rempe_2002, FeedbackTheory_Steck_2004, FeedbackTheory_Steck_2006} to compute the control signal. 
Unmodeled nonlinearities, parameter drift, and environmental noise can degrade estimation fidelity and destabilize feedback.

To overcome these limitations, it is natural to seek control strategies that can improve directly from experimental data. Machine learning provides a general framework for navigating such high-dimensional, nonlinear parameter spaces, and has already proven effective as an optimizer at the frontier of quantum controls. Notable examples include automated optimization of ultracold atomic cloud production~\cite{MLOOP_Hush_2016, MLOOP_Vendeiro_2022, MLOOP_Peters_2024} and
quantum error correction~\cite{QECRL_Weiss_2018}. 
Among these methods, deep reinforcement learning (RL), which learns control policies from interaction with an environment, is therefore particularly well-suited when intermediate observations are available. While recent work has applied RL to control bulk properties of cold-atom ensembles in magneto-optical traps~\cite{AtomRL_Reinschmidt_2024, AtomRL_Leblanc_2023}, extending RL to quantum-noise-limited feedback control of a single atom constitutes a qualitatively new frontier in atomic, molecular, and optical physics.

Here, we demonstrate the first feedback cooling of a single neutral atom in a high-finesse optical cavity using deep reinforcement learning. By continuously observing the cavity transmission, which is dispersively perturbed by the atom's position within the cavity mode, our agent learns to modulate the optical trapping potential to remove atomic kinetic energy. We begin with simulated training on a numerical environment before deploying the agent on the experimental hardware, where the RL agent learns a robust, nonlinear control policy that actively damps the atom's motion in a quantum-noise-limited environment. We find that the learned policy cools significantly faster than a standard differentiator controller \cite{ExternalFeedback_Vuletic_2007} while maintaining comparable atom retention, and we later trace this advantage to a learned force profile with a twice-larger peak damping force.

\section*{Experimental Apparatus}

\begin{figure*}
    \centering
    \includegraphics[width=1\textwidth]{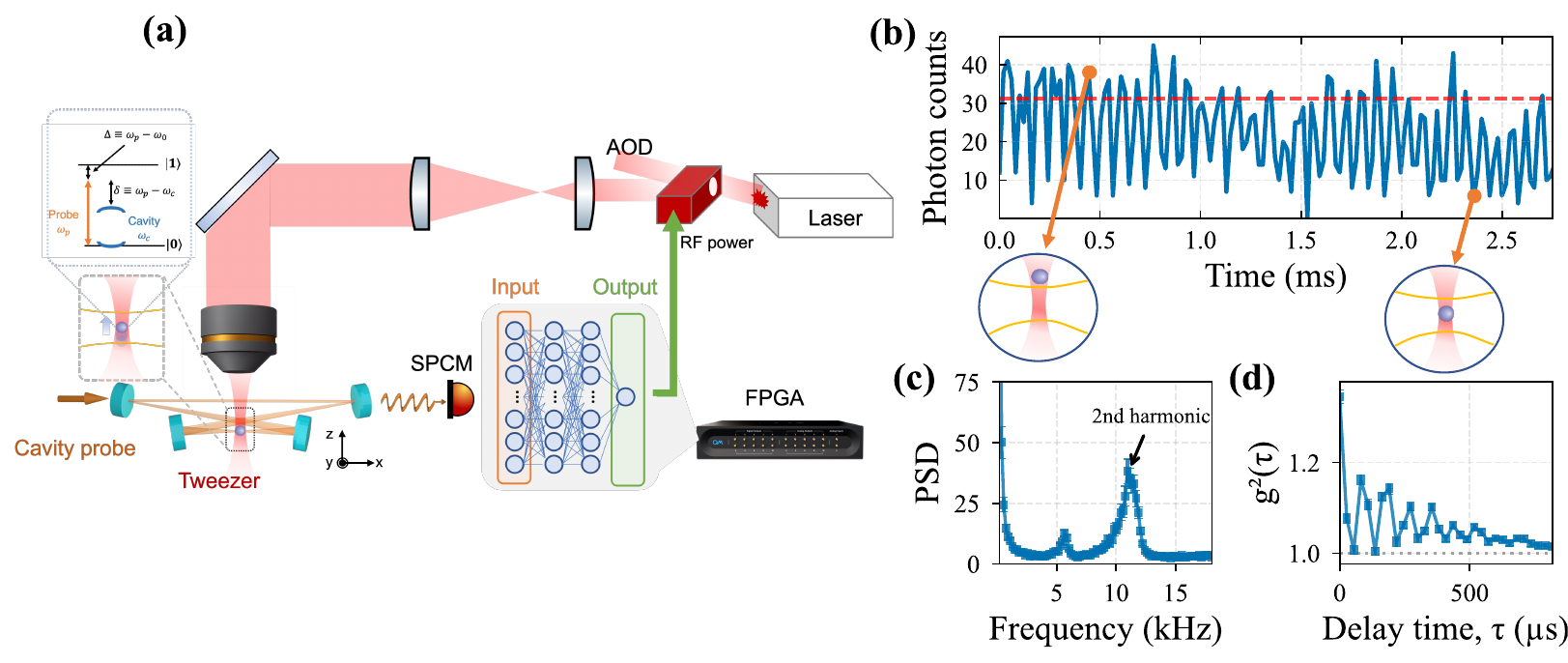} 
    \caption{\textbf{Observing single-atom motion in an optical cavity.} \textbf{a}, Experimental schematic. A single Cs atom is trapped in a 937-nm optical tweezer and placed at the center of a high-finesse bow-tie cavity. The cavity transmission is monitored with a single-photon counting module (SPCM) to track the atom's trajectory. \textbf{b}, A typical real-time trajectory of a single atom, showing oscillations in the transmitted photon counts. The signal maxima (minima) correspond to the atom's passage through the edge (center) of the cavity mode. The dashed red line indicates the empty-cavity average transmission. \textbf{c}, The power spectral density (PSD) of the transmission signal, averaged over 143 trajectories, reveals a peak at \SI{11.16\pm0.03}{kHz}, corresponding to twice the longitudinal trap frequency. We attribute the broad width to the anharmonicity from the Gaussian potential explored by hotter atoms. \textbf{d}, The second-order correlation function of the transmitted photons, $g^{(2)}(\tau)$, averaged over 143 trajectories, shows decaying oscillations that indicate a motional coherence time of several hundred \SI{}{\micro s}. For the above measurements, no feedback is applied. 
    }
    \label{fig:fig1}
\end{figure*}

Our goal is to cool the motional degree of freedom of a single trapped atom by inferring its position and using real-time modulation of the trapping potential to remove kinetic energy. The key physical mechanism we utilize to measure atomic position is dispersive cavity QED readout. In this regime, the atom functions analogously to a moving dielectric particle with a position-dependent refractive index. When the atom moves into the center of the cavity mode, it shifts the cavity's resonant frequency. Because we probe the cavity at a fixed frequency, this atomic-induced shift causes a real-time change in the transmitted light intensity proportional to the square of the local atom--cavity coupling strength $g(\vec{r})$ and the atomic motion is continuously imprinted onto the cavity transmission signal.

Our experimental setup, shown schematically in Fig.~\ref{fig:fig1}(a), is designed to leverage this position-to-photon-count transduction. A single caesium atom is confined in a 937-nm optical tweezer (1/$e^2$ waist \SI{1.52\pm0.02}{\micro m}) and positioned at the center of the $\mathrm{TEM}_{00}$ mode of the optical bow-tie cavity, which has a waist of $w_c=\SI{7}{\micro m}$. By applying a bias magnetic field (\SI{4.8}{G}) along the cavity mode propagation direction ($x$-axis), we isolate the cycling transition between the ground state $\ket{6^2S_{1/2}, F=4,m_F=4} \equiv\ket{0}$ and the excited state $\ket{6^2P_{3/2}, F=5,m_F=5}\equiv\ket{1}$ to effectively create a two-level system. The tweezer's weakly-confining longitudinal axis is aligned transverse (z-axis) to the cavity mode, making the transmission signal highly sensitive to the atom's motional excursions along this direction.

\subsection*{Direct Observation of Atomic Motion}

To directly resolve this motion in real time, we probe the cavity with a weak $\sigma^+$-polarized laser of frequency $\omega_p$ that is detuned from the atomic resonance $\omega_0$ by $\Delta=\omega_p-\omega_0=2\pi\times \SI{25}{MHz}$ while remaining on the empty-cavity resonance $\omega_c$ ($\delta=\omega_p-\omega_c=0$). Figure~\ref{fig:fig1}(b) shows a typical time trace of the transmitted photon counts, where the oscillations arise because the atom samples regions of different coupling strengths as it moves along the longitudinal axis of the tweezer. When the atom passes through the cavity center and couples most strongly, it induces the largest dispersive cavity shift and the transmission is most strongly suppressed; near the mode edges the coupling is weaker, the induced shift is smaller, and the transmission increases. We therefore attribute the valleys (peaks) in the signal to the atom being near the center (turning points) of its trajectory within the cavity mode. Over an ensemble of data, the transmission exhibits a dominant frequency component at \SI{11.16\pm0.03}{kHz} in Fig.~\ref{fig:fig1}(c), which corresponds to twice the longitudinal trap frequency (due to the symmetric shape of the cavity mode). Finally, Fig.~\ref{fig:fig1}(d) shows the second-order correlation function of the detected transmitted photons, $g^{(2)}(\tau)$, with decaying oscillations that indicate motional coherence over several hundred microseconds.

\section*{Reinforcement Learning for Real-Time Feedback}

We formulate the system above as a feedback-cooling control task and train a Soft Actor-Critic (SAC) reinforcement learning agent \cite{SAC_Haarnoja_2018} to reduce the atom’s kinetic energy. The agent is trained to observe the recent history of cavity transmission and apply continuous trap-depth updates to minimize motional energy while avoiding atom loss. For our control policy, we utilize a multi-layer perceptron (MLP), a type of feedforward neural network composed of sequential layers of linear transformations and nonlinear activation functions. An MLP can be viewed as an arbitrary function approximator, mapping input variables to outputs by constructing nonlinear combinations of the inputs, analogous to building a basis expansion with tunable coefficients.  

Training an RL agent directly on a cold atom experiment may be slow or impractical due to the large number of iterations required and the slow cycle times ($\gtrsim \SI{1}{s}$ per shot). To circumvent this challenge, we adopt a \emph{sim-to-real} transfer learning approach~\cite{SimReal_Jakobi_1995, SimReal_Andrei_2018, SimReal_Wenshuai_2020, 
AtomRL_Reinschmidt_2024}, wherein an agent is first trained on a numerical simulation based on the stochastic equations of motion for the coupled atom-cavity system before transferring the knowledge to the experiment. The experimental agent control policy, hereafter referred to as MLP (Expt.), is trained on the experimental hardware and initialized with a replay buffer seeded by the simulation-trained control policy (see Methods), hereafter referred to as MLP (Sim.). Full state, action, reward definitions, environment, network architecture, and training protocol are provided in Methods. 

\begin{figure*}
\centering
\includegraphics[width=1.\textwidth]{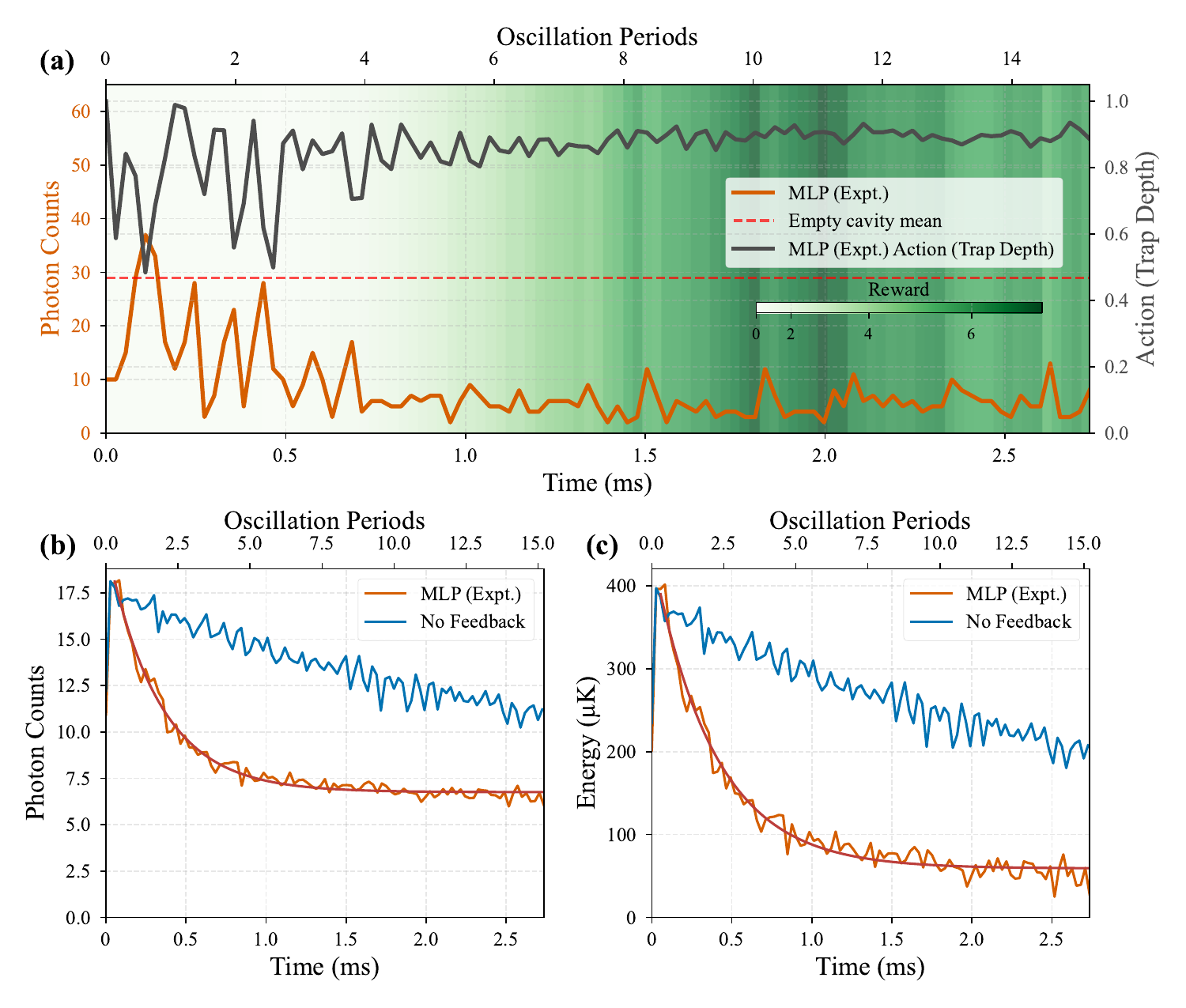}
\caption{\textbf{Demonstration of Feedback Cooling with a Reinforcement Learning Agent.} \textbf{a}, A single experimental trajectory showing the real-time relationship between the observed cavity transmission (orange) and the feedback action (grey) applied by the experimentally-trained MLP. The agent responds to the initial high-amplitude oscillations, which indicate high energy, by modulating the trap potential to actively damp the atom's motion. The reward the agent receives at each timestep is indicated in green. \textbf{b}, Averaged photon counts over many experimental runs, which contrast the performance of the RL agent (orange) with the passively cooled case (blue). Active feedback rapidly drives the signal toward the low-energy transmission level. \textbf{c}, The averaged photon count data from (b) is converted to longitudinal energy using a model of cavity transmission. For trajectories conditioned on end-of-sequence survival, an exponential fit to the energy decay for the actively cooled case (orange) yields a $1/e$ cooling time of $\SI{388\pm14}{\micro s}$, corresponding to a cooling timescale of just $\sim 2$ longitudinal motional periods.} 
\label{fig:fig2}
\end{figure*}

We find that the resulting experimentally-trained multi-layer perceptron demonstrates effective feedback cooling, as shown by the typical cooling trajectory in Fig.~\ref{fig:fig2}(a). Initially, observable oscillations in the transmitted photon counts (orange) indicate a large atomic energy, and the agent applies a sequence of carefully timed actions (grey) to modulate the trap potential to systematically remove this energy. By averaging over 148 experimental runs and post-selecting on the atom remaining trapped at the end of the sequence so that the mean is not biased by early loss (Fig.~\ref{fig:fig2}(b,c)), we extract a characteristic $1/e$ cooling time of $\SI{388\pm14}{\micro s}$. This corresponds to just $\sim2$ periods of the atom's oscillation in the trap, and the atom is cooled to a longitudinal temperature of $T_z = \SI{45 \pm 10}{\micro K}$ from an initial average axial energy of \SI{429\pm48}{\micro K} (see Methods for temperature calculation details). For comparison, with identical probing but no trap modulation (blue), we observe only slow passive cooling from cavity dynamical backaction (see Supplementary Information); the much faster decay with feedback confirms that real-time control dominates in our regime.

\subsection*{Experimental Validation and Performance}

\begin{figure*}
    \centering
    \includegraphics[width=1.0\textwidth]{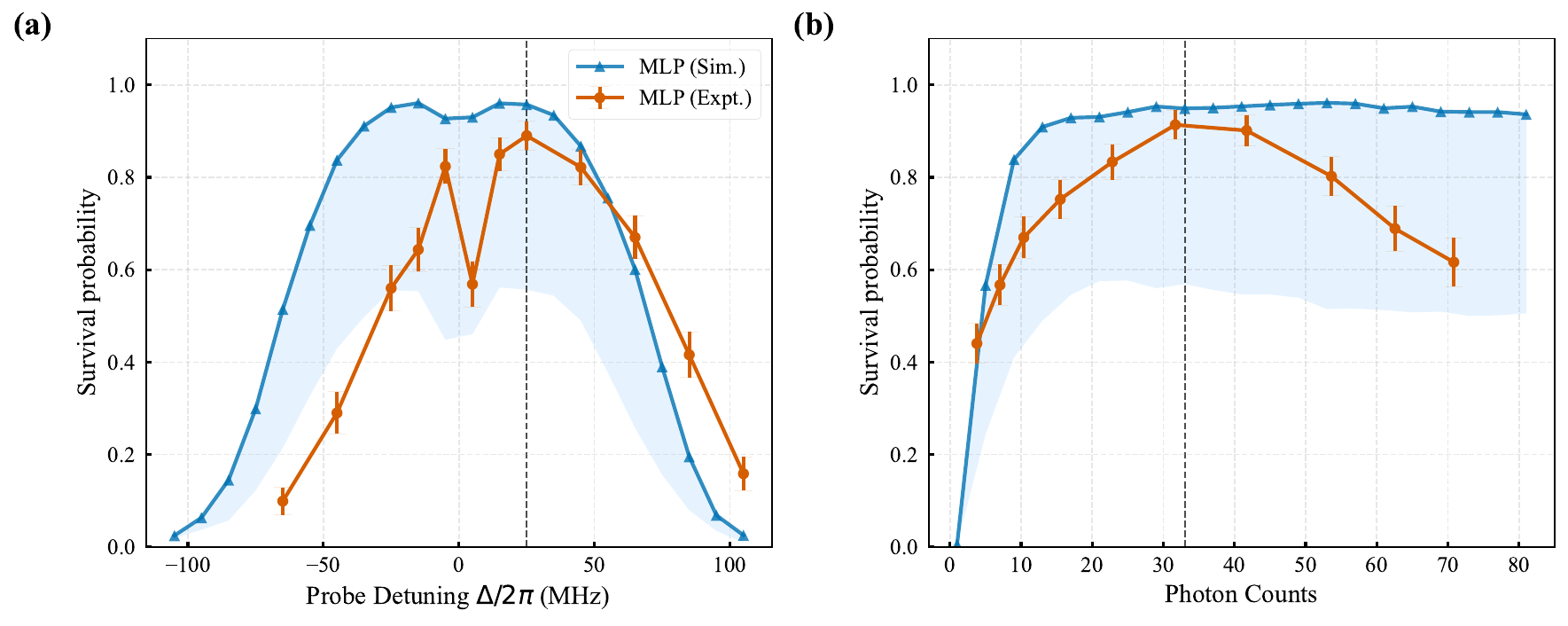}

        \caption{\textbf{ \emph{Sim-to-Real} Experimental Validation.}   
        \textbf{a, b}, Comparison of simulated and experimental survival trends for the experimentally-trained MLP (Expt., orange) and the simulation-trained MLP (Sim., blue). Performance is measured across a range of probe detunings (\textbf{a}) and incident detected photon counts within a time bin (\textbf{b}). Error bars denote the standard error of the mean. The vertical lines indicate the parameters at which the agents were trained. The shaded blue region captures the systematic uncertainty of the simulation.}
\label{fig:fig3}
\end{figure*}

To assess whether the simulation captures the relevant experimental trends near our operating point, we compare an RL-agent control policy trained on experiment against one trained in simulation. Because our measurement record primarily reflects longitudinal motion, a controller can inadvertently heat unobserved radial degrees of freedom when modulating the trap. We therefore use end-of-sequence survival as the central experimental metric: it penalizes overly aggressive cooling and is directly optimized during training because atom loss terminates an episode and removes future reward.

We compare the experimentally-measured survival probability of MLP (Expt.) against the performance of MLP (Sim.) in simulation while varying the atom-probe detuning, $\Delta$, and the empty-cavity photon rate (Fig.~\ref{fig:fig3}(a,b)). The shaded region in the figure shows the simulation's systematic uncertainty, accounting for factors including shot-to-shot photon count fluctuations and potential tweezer-cavity misalignments. The simulation captures the qualitative survival trends near our operating point $\Delta=2\pi\times\SI{25}{MHz}$, particularly on the blue-detuned side of the atomic resonance ($\Delta >0$), leading us to conclude that the simulation serves as a useful training model in this regime. As the probe is tuned to the red side (lower frequency), stronger disagreement emerges, which we primarily attribute to contributions from further red-detuned excited states that violate our idealized two-level model, as well as the internal magnetic structure of the atom (Zeeman sublevels), which are neglected in the simulation. 
Additional sources of discrepancy may include a non-thermal initial energy distribution, unmodeled experimental noise, and dipole force fluctuations, which arise if the optical tweezer wavelength is not perfectly `magic' (a condition where the trapping potential is identical for both the ground and excited electronic states). At high photon counts (Fig.~\ref{fig:fig3}(b)), we hypothesize that the observed reduction in experimental survival is similarly due to increased heating from these unmodeled multi-level effects. Importantly, even though these effects are not modeled in the simulation, our experimental RL agent still learns to adapt to these unmodeled dynamics to effectively cool.

\subsection*{Agent Benchmarking Comparison}

\begin{figure*}
    \centering
    \includegraphics[width=0.83\textwidth]{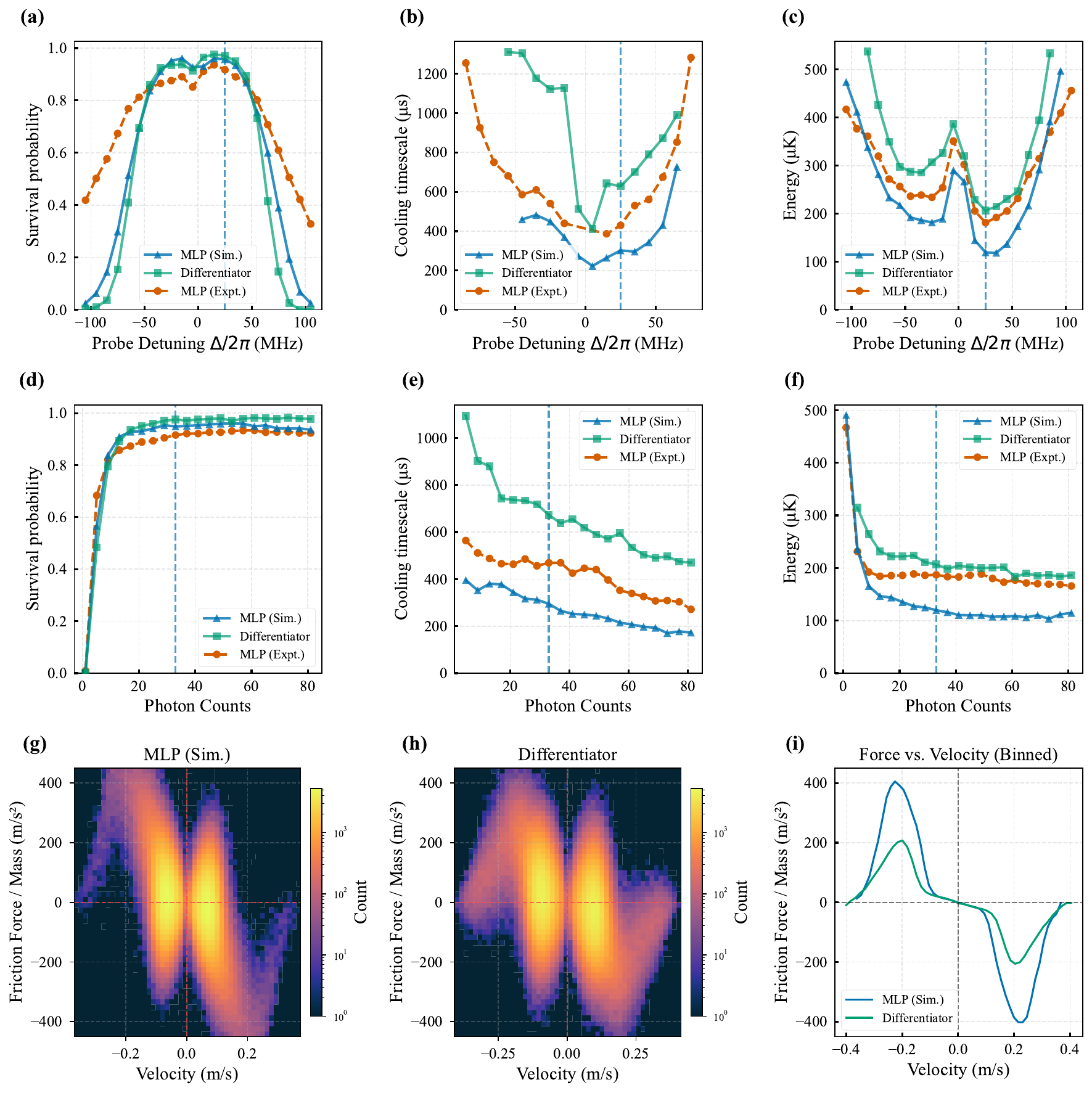}
        \caption{\textbf{Benchmarking of Control Policies.}  \textbf{a–f}, Systematic benchmarking of three distinct feedback controllers in simulation: one simulation-trained MLP (Sim.), the experimentally-trained MLP (Expt.), and a differentiator-based controller. Each data point is generated from 4,000 simulation runs with the respective model. Performance is evaluated across the parameter space of probe detuning (top row) and photon count (bottom row) based on three metrics: survival probability (\textbf{a, d}), cooling timescale (\textbf{b, e}), and final energy after 20 feedback steps (\textbf{c, f}). Notably, the experimentally-trained policy demonstrates strong and robust performance when evaluated in the simulated environment. The dashed vertical lines indicate the parameters at which the agents were trained. \textbf{g–i}, A detailed comparison of the cooling dynamics and effective force profiles for the MLP (Sim.) agent and the differentiator controller, evaluated over 40,000 simulated episodes. \textbf{g, h}, Phase-space distributions of the effective friction force per unit mass (acceleration) applied by the MLP (Sim.) controller (\textbf{g}) and the differentiator (\textbf{h}). The MLP (Sim.) controller applies a sharper and better-defined force profile, consistent with greater robustness to measurement noise. \textbf{i}, The averaged effective force as a function of velocity, showing the MLP agent learns to apply a peak damping force approximately twice as strong as the differentiator.}
    \label{fig:fig4}
\end{figure*}

We benchmark the performance of learned policies against a differentiator control strategy~\cite{ExternalFeedback_Vuletic_2007} across a two-dimensional parameter space to gain insight into how the nonlinear policy exploits complex, potentially unmodeled system dynamics and adapts to changing experimental conditions that a simpler linear model cannot account for (Fig.~\ref{fig:fig4}(a-f)). We evaluate three distinct controllers within our simulation: a simulation-trained MLP optimized for fast cooling hereafter referred to as MLP (Sim.), the experimentally-trained MLP (Expt.), and a differentiator controller tuned to closely match the survival probability of the MLP (Sim.) agent at the training point.  All models were tuned at an empty-cavity photon count of 33 (chosen to balance heating and SNR) and $\Delta = 2\pi\times\SI{+25}{MHz}$ (chosen so that an atom at the center of the cavity mode at low temperature corresponds to a high-slope region on the transmission Lorentzian).

We assess each controller's performance based on three key performance metrics: atom survival probability, cooling timescale, and the final energy achieved after 20 feedback steps. For the cooling time metric in Fig.~\ref{fig:fig4}(b,e), we assign a cooling timescale only when the averaged energy trace can be meaningfully fit: the fit includes only time points with at least 200 surviving episodes, and each point is weighted by the standard error of the mean energy at each time point. Fits with poor exponential goodness-of-fit or unconstrained timescale uncertainty were excluded from the cooling time metric. To distill these results into a direct comparison, we introduce a composite performance metric that quantifies both the efficacy and robustness of each policy (Table~\ref{tab:performance_metrics}). For each parameter sweep (detuning and photon count) and for each of the three performance metrics, we compute a score as the geometric mean of two normalized factors: \textit{peak efficacy}, the agent's best-achieved result relative to the best performance across all agents, and \textit{robustness}, the width of the parameter range over which the controller remains near its optimum, normalized to the largest such width across all agents. For survival, this interval is defined by performance remaining above 50\% of the peak value; for 20-step energy and cooling timescale, it is defined by the metric remaining below 1.5 times its optimum value. The Combined row reports the geometric mean of the corresponding Detuning and Photon Count scores, while the Overall column reports the geometric mean of the Survival, Cooling, and Energy entries within a given row. The MLP (Sim.) agent achieves a combined average score of 0.898 over photon count and detunings, outperforming the differentiator-based controller. The experimentally-trained agent, MLP (Expt.), also performs remarkably well with a combined score of 0.804. This emergent robustness is a key finding of our work, demonstrating that fine-tuning on the physical hardware allows the agent to discover policies that are resilient to a broader range of operating conditions. We attribute this to its exposure to real-world experimental noise and unmodeled dynamics imperfectly captured in simulation.

To understand the physical origin of the MLP agent's superior performance, we analyze its learned control strategy in simulation and compare it directly to the differentiator (Fig.~\ref{fig:fig4}(g-i)). The advantage is immediately apparent in the phase-space distributions of the applied feedback force. The differentiator controller (Fig.~\ref{fig:fig4}(h)) exhibits a broad, diffuse distribution, indicating that the applied force for a given velocity is poorly defined due to measurement noise or nonlinearities. In contrast, the MLP (Sim.) controller (Fig.~\ref{fig:fig4}(g)) learns a much sharper, more deterministic force profile. This demonstrates the agent's more sophisticated nonlinear control policy that maps a richer history of observations to a feedback action. As a result, the MLP is able to infer the atom’s underlying state with higher fidelity than the differentiator. This superior state estimation allows the agent to apply a peak damping force that is approximately twice as large as that of the differentiator controller without sacrificing stability (Fig.~\ref{fig:fig4}(i)), enabling both faster and more effective cooling.

\begin{table}[h]
\centering
\scriptsize

\begin{tabular}{@{}llcccc@{}}
\toprule
\textbf{Metric Group} & \textbf{Model} & \textbf{Survival} & \textbf{Cooling} & \textbf{Energy} & \textbf{Overall} \\
\midrule
\multirow{3}{*}{\textbf{Detuning}} & MLP (Sim.) & 0.846 & 0.839 & 0.871 & 0.852 \\
 & MLP (Expt.) & 0.979 & 0.759 & 0.790 & 0.837 \\
 & Differentiator & 0.814 & 0.380 & 0.757 & 0.616 \\
\midrule
\multirow{3}{*}{\textbf{Photon Count}} & MLP (Sim.) & 0.984 & 0.899 & 0.957 & 0.946 \\
 & MLP (Expt.) & 0.975 & 0.598 & 0.790 & 0.772 \\
 & Differentiator & 0.991 & 0.602 & 0.734 & 0.759 \\
\midrule
\multirow{3}{*}{\textbf{Combined}} & MLP (Sim.) & 0.913 & 0.869 & 0.913 & 0.898 \\
 & MLP (Expt.) & 0.977 & 0.674 & 0.790 & 0.804 \\
 & Differentiator & 0.898 & 0.478 & 0.746 & 0.684 \\
\bottomrule

\end{tabular}

\caption{\textbf{Quantitative Performance Metrics.} A summary of the composite performance metrics for each feedback controller that provide a quantitative analysis of the data in Fig.~\ref{fig:fig4}(a-f). For each sweep and metric, the score is the geometric mean of normalized peak efficacy and normalized robustness width. For survival, robustness is the parameter interval over which performance remains above 50\% of its peak value; for 20-step energy and cooling timescale, it is the interval over which the metric remains below 1.5 times its optimum value. The Combined row is the geometric mean of the Detuning and Photon Count scores, and the Overall column is the geometric mean across Survival, Cooling, and Energy.}
\label{tab:performance_metrics}
\end{table}

\section*{Outlook}

The techniques presented here open several avenues for future research at the intersection of quantum control and machine learning in complex quantum systems. For instance, the agent’s action space could be readily expanded to include simultaneous control over additional experimental parameters, such as the timing and profile of the trap modulation or the detuning of the cavity probe $\delta$. Such multi-dimensional optimization may be intractable for traditional model-based controllers but is a natural extension for an RL agent. Similarly, enhancing the agent’s observational capabilities by providing access to complementary information, such as the phase of the transmitted light, could unlock even more sophisticated control strategies. For systems where the quantum dynamics become too complex to simulate, our demonstrated technique enables a powerful ``classical-to-quantum'' transfer paradigm, where an agent is trained in a tractable classical model and then refined directly on the real quantum hardware when accurate simulation is no longer feasible. This may provide a pathway to discover control protocols for otherwise difficult-to-model control problems, such as navigating the complex energy landscapes of unresolved cavity sideband cooling~\cite{Sidebandcool_Wang_2025}.

Looking forward, a compelling direction is the integration of our control methods with next-generation quantum hardware. For example, recently developed low-finesse micro-cavity arrays offer a platform for scalable, parallel atom-photon interfacing~\cite{ParallelCavityArray_Simon_2025}. 
Such platforms will require precisely the kind of autonomous, robust, and scalable control demonstrated in this work. Applying our framework to such small-waist systems would naturally overcome many of the limitations of our current setup; with sensitivity to all three motional degrees of freedom, an RL agent could learn to cool in three dimensions and retain the atom at low temperatures without the radial escape mechanism present in our work. If the observed cooling timescale of just a few oscillation periods holds for the higher trapping frequencies in these systems, cooling speeds on the order of tens of microseconds could be achieved. Additionally, the temperature reached in our system is largely limited by atomic positional sensitivity in the \SI{7}{\micro m} cavity waist. Micron-scale cavity waists~\cite{ParallelCavityArray_Simon_2025} would enable larger sensitivities to position displacements and cooling to lower temperatures. Furthermore, the underlying principles of this work are broadly applicable beyond single atoms, extending naturally to the feedback cooling of levitated nanoparticles~\cite{SubwavelengthCooling_Diehl_2018, ColdDamping_Tebbenjohanns_2019, DigitalParametricFeedback_Zheng_2019, FeedbackNanoparticle_Aikawa_2021} and potentially to molecules~\cite{ExternalFeedback_Vuletic_2007}.

These results point to a practical route for applying learning-based feedback control in quantum experiments where the relevant state is only partially observed and accurate analytical models are incomplete. By combining physics-based simulation with limited hardware fine-tuning, the approach can exploit existing physical understanding while adapting to experimental imperfections that are difficult to model. Extending this strategy to multi-parameter control, additional measurement channels, and platforms with sensitivity to all motional degrees of freedom could enable faster cooling, improved state preparation, and more robust operation of cavity-coupled atom arrays and other individually controlled quantum systems.

\section*{Methods}

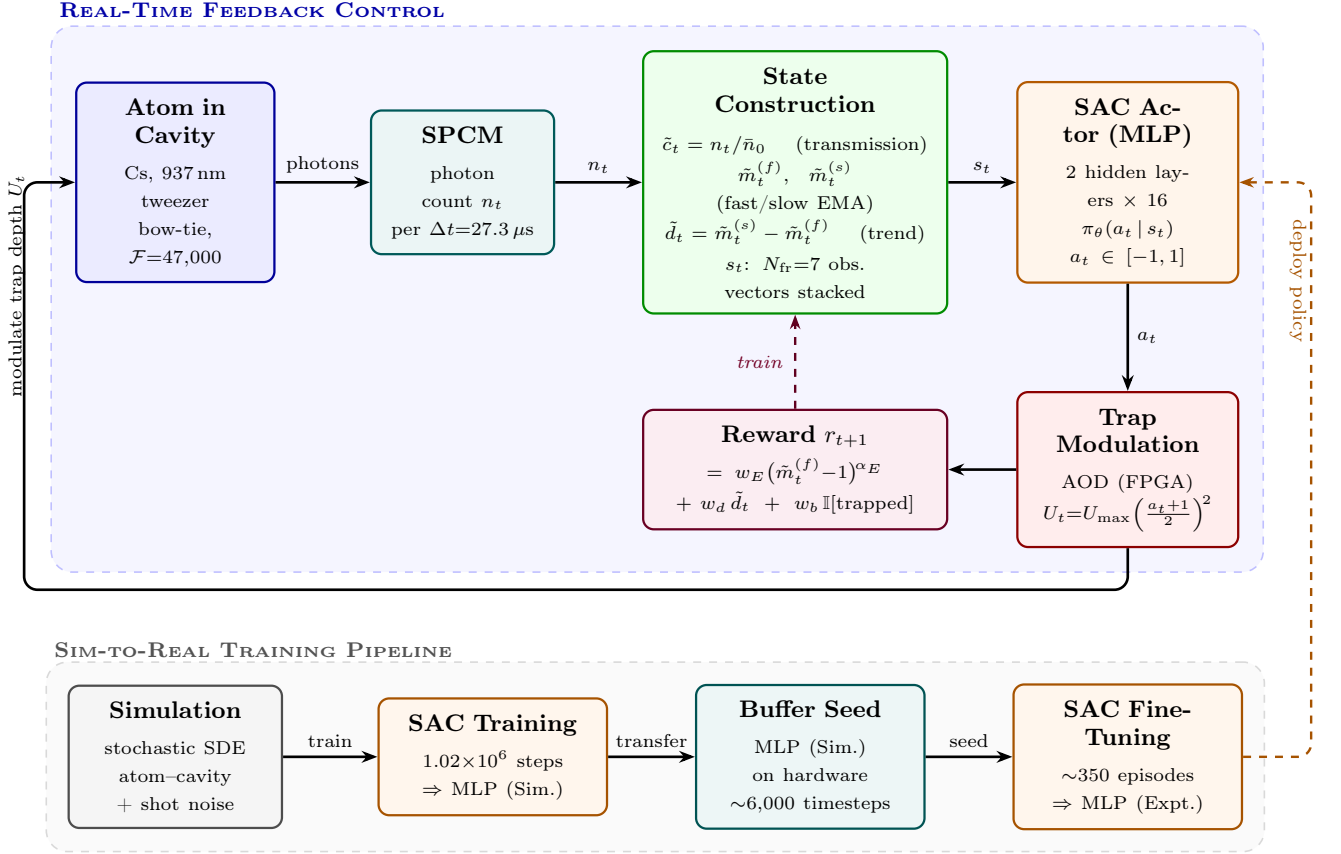
\begin{figure*}
\centering
\begin{tikzpicture}[
  B/.style 2 args={rectangle, rounded corners=4pt,
                   draw=#1, fill=#2,
                   line width=0.85pt, align=center, inner sep=6pt},
  arr/.style={-{Stealth[length=6pt,width=4.5pt]}, line width=1pt},
  darr/.style={-{Stealth[length=5.5pt,width=4pt]},
               line width=0.9pt, dashed},
  font=\small,
]


\node[B={blue!60!black}{blue!8},
      text width=2.2cm, minimum height=1.6cm] (atom) at (0,0)
  {\textbf{Atom in Cavity}\\[3pt]
   {\scriptsize Cs, 937\,nm tweezer}\\
   {\scriptsize bow-tie, $\mathcal{F}{=}47{,}000$}};

\node[B={teal!70!black}{teal!8},
      text width=2.0cm, minimum height=1.6cm] (spcm) at (3.8,0)
  {\textbf{SPCM}\\[3pt]
   {\scriptsize photon count $n_t$}\\
   {\scriptsize per $\Delta t{=}27.3\,\mu$s}};

\node[B={green!55!black}{green!7},
      text width=3.6cm, minimum height=2.1cm] (state) at (8.2,0)
  {\textbf{State \\ Construction}\\[4pt]
   {\scriptsize $\tilde{c}_t = n_t/\bar{n}_0$ \quad (transmission)}\\[1pt]
   {\scriptsize $\tilde{m}^{(f)}_t,\;\tilde{m}^{(s)}_t$\quad (fast/slow EMA)}\\[1pt]
   {\scriptsize $\tilde{d}_t = \tilde{m}^{(s)}_t-\tilde{m}^{(f)}_t$ \quad (trend)}\\[1pt]
   {\scriptsize $s_t$: $N_\text{fr}{=}7$ obs.\ vectors stacked}};

\node[B={orange!70!black}{orange!8},
      text width=2.5cm, minimum height=1.6cm] (actor) at (12.6,0)
  {\textbf{SAC Actor (MLP)}\\[3pt]
   {\scriptsize 2 hidden layers $\times$ 16}\\
   {\scriptsize $\pi_\theta(a_t\,|\,s_t)$}\\
   {\scriptsize $a_t \in [-1,1]$}};

\node[B={red!55!black}{red!7},
      text width=2.5cm, minimum height=1.5cm] (aod) at (12.6,-3.8)
  {\textbf{Trap \\ Modulation}\\[3pt]
   {\scriptsize AOD (FPGA)}\\
   {\scriptsize $U_t{=}U_{\max}\!\left(\frac{a_t{+}1}{2}\right)^{\!2}$}};

\node[B={purple!55!black}{purple!7},
      text width=3.6cm, minimum height=1.5cm] (rew) at (8.2,-3.8)
  {\textbf{Reward} $r_{t+1}$\\[3pt]
   {\scriptsize $= w_E\bigl(\tilde{m}^{(f)}_t{-}1\bigr)^{\!\alpha_E}$}\\[1pt]
   {\scriptsize $+\;w_d\,\tilde{d}_t \;+\; w_b\,\mathbb{I}[\mathrm{trapped}]$}};

\draw[arr] (atom)  -- (spcm)
    node[midway, above, font=\scriptsize] {photons};
\draw[arr] (spcm)  -- (state)
    node[midway, above, font=\scriptsize] {$n_t$};
\draw[arr] (state) -- (actor)
    node[midway, above, font=\scriptsize] {$s_t$};
\draw[arr] (actor) -- (aod)
    node[midway, right, font=\scriptsize] {$a_t$};
\draw[arr] (aod.west) -- (rew.east);

\coordinate (Rdown) at ($(aod.south)+(0,-0.55)$);
\coordinate (Rleft) at ($(-2.0,0 |- Rdown)$);
\coordinate (Rup)   at ($(-2.0,0)$);
\draw[arr, rounded corners=5pt]
    (aod.south) -- (Rdown) -- (Rleft) -- (Rup) -- (atom.west)
    node[font=\scriptsize, pos=0.42, below, rotate=90, xshift=-10ex, yshift=4.5ex] {modulate trap depth $U_t$};

\draw[darr, draw=purple!50!black]
    (rew.north) -- (state.south)
    node[midway, left=0.05cm, font=\scriptsize\itshape,
         text=purple!60!black] {train};

\begin{scope}[on background layer]
  \node[fill=blue!4, rounded corners=8pt,
        draw=blue!25, dashed, line width=0.6pt,
        fit=(atom)(spcm)(state)(actor)(aod)(rew),
        inner sep=9pt] (fbbox) {};
\end{scope}
\node[font=\footnotesize\bfseries, text=blue!65!black,
      anchor=south west] at (fbbox.north west)
    {\textsc{Real-Time Feedback Control}};


\node[B={gray!60!black}{gray!8},
      text width=2.4cm, minimum height=1.4cm] (sim) at (0,-7.6)
  {\textbf{Simulation}\\[3pt]
   {\scriptsize stochastic SDE}\\
   {\scriptsize atom--cavity + shot noise}};

\node[B={orange!65!black}{orange!7},
      text width=2.6cm, minimum height=1.4cm] (simtr) at (4.2,-7.6)
  {\textbf{SAC Training}\\[3pt]
   {\scriptsize $1.02{\times}10^6$ steps}\\
   {\scriptsize $\Rightarrow$ MLP (Sim.)}};

\node[B={teal!65!black}{teal!7},
      text width=2.6cm, minimum height=1.4cm] (buf) at (8.4,-7.6)
  {\textbf{Buffer Seed}\\[3pt]
   {\scriptsize MLP (Sim.) on hardware}\\
   {\scriptsize ${\sim}6{,}000$ timesteps}};

\node[B={orange!65!black}{orange!7},
      text width=2.6cm, minimum height=1.4cm] (expt) at (12.6,-7.6)
  {\textbf{SAC Fine-Tuning}\\[3pt]
   {\scriptsize ${\sim}350$ episodes}\\
   {\scriptsize $\Rightarrow$ MLP (Expt.)}};

\draw[arr] (sim)   -- (simtr)
    node[midway, above, font=\scriptsize] {train};
\draw[arr] (simtr) -- (buf)
    node[midway, above, font=\scriptsize] {transfer};
\draw[arr] (buf)   -- (expt)
    node[midway, above, font=\scriptsize] {seed};

\coordinate (D1) at ($(expt.east)+(0.9,0)$);
\coordinate (D2) at ($(D1)+(0,7.60)$);
\draw[darr, draw=orange!65!black, rounded corners=5pt]
    (expt.east) -- (D1) -- (D2) -- ([yshift=0ex]actor.east)
    node[font=\scriptsize, pos=0.57, right=0.6cm, below=1.3cm, rotate=-90,
         text=orange!65!black] {deploy policy};

\begin{scope}[on background layer]
  \node[fill=gray!4, rounded corners=8pt,
        draw=gray!30, dashed, line width=0.6pt,
        fit=(sim)(simtr)(buf)(expt),
        inner sep=8pt] (trbox) {};
\end{scope}
\node[font=\footnotesize\bfseries, text=gray!65!black, yshift=-0.4ex,
      anchor=south west] at (trbox.north west)
    {\textsc{Sim-to-Real Training Pipeline}};

\end{tikzpicture}
\caption{\textbf{Overview of the RL feedback cooling method.}
  (\textit{Top}) Real-time control loop: the SPCM records cavity transmission photon counts
  $n_t$; engineered features (normalized transmission $\tilde{c}_t$, fast/slow exponential
  moving averages $\tilde{m}^{(f,s)}_t$, and trend $\tilde{d}_t$) are stacked into a state $s_t$
  and fed to the SAC actor, which outputs an action $a_t$ that sets the optical tweezer trap depth
  via the AOD.  The reward $r_{t+1}$ incentivizes rapid energy removal while preserving atom
  confinement.  (\textit{Bottom}) Sim-to-real training pipeline: the SAC agent is first trained
  entirely in a stochastic SDE simulation (1.02\,$\times\,10^6$ steps), producing MLP (Sim.);
  this policy is then used to seed the replay buffer (${\sim}6{,}000$ steps collected on hardware),
  enabling rapid fine-tuning on the physical apparatus in ${\sim}350$ episodes to produce the
  experimentally-deployed MLP (Expt.).}
\label{fig:methods}
\end{figure*}

An overview of the complete method---including the real-time feedback control loop and the sim-to-real training pipeline---is shown schematically in Fig.~\ref{fig:methods}.


\subsection*{Atom Preparation and Cavity Parameters}
We prepare a single-atom system by first collecting caesium atoms in a magneto-optical trap, after which we apply \SI{35}{ms} of 3D optical molasses. This cooling phase induces light-assisted collisions~\cite{LightAssistedCollisions_Gomer_1999, LightAssistedCollisions_Julienne_1999, LightAssistedCollisions_Grangier_2001}, 
resulting in the probabilistic loading of a single atom into a 937-nm magic-wavelength optical tweezer~\cite{Magic_Kim_2003, Magic_Ye_1999}. The trap is generated by an acousto-optic deflector and has a power of \SI{10.0\pm0.5}{mW} and a $1/e^2$ waist of \SI{1.52\pm0.02}{\micro m}, yielding a calculated trap depth of $U_{\textrm{max}}=\SI{37\pm2}{MHz}$ and a radial trapping frequency of $2\pi\times\SI{71\pm3}{kHz}$. After confirming the presence of an atom using projective imaging with a 1D retroreflected molasses beam, we characterized its motional state via release and recapture.

The single caesium atom was positioned at the waist of a four-mirror, optical bow-tie cavity with a finesse of $\mathcal{F}=47,000$. The cavity mode had a Gaussian waist of $w_c=\SI{7}{\micro m}$ and a Rayleigh length of $x_c=\SI{180}{\micro m}$~\cite{Adaptive_Peters_2024}. A bias magnetic field of \SI{4.8}{G} was applied along the cavity axis ($x$-axis). The atom functions as a two-level system defined by the cycling transition between the ground state $\ket{6^2S_{1/2}, F=4,m_F=4} \equiv\ket{0}$ and the excited state $\ket{6^2P_{3/2}, F=5,m_F=5}\equiv\ket{1}$. A near-$\sigma^+$ polarized probe light optically pumped the atom into the stretched state $\ket{F=4,m_F=4}$, where it coupled maximally to the cavity mode. The relevant cavity QED parameters are $(g_{\mathrm{max}}, \kappa,\Gamma)=2\pi\times(\SI{1.03\pm0.04}{},\SI{0.040\pm0.003}{},\SI{5.234}{})$~MHz, where $g_{\mathrm{max}}$ is the maximum single-photon Rabi frequency at the center of the cavity mode, $\kappa$ is the cavity FWHM linewidth, and $\Gamma$ is the caesium $D_2$-line FWHM linewidth. The position-dependent coupling strength is given by:
\begin{equation}
\label{eqn:coupling}
    g(y, z)\approx g_{\mathrm{max}}e^{-\frac{(y^2+z^2)}{w_c^2}}
\end{equation}
Atomic motion is transduced into fluctuations in the cavity transmission, monitored by a single-photon counting module (SPCM) with an estimated total detection efficiency of $\sim12.5\% $ in the apparatus (accounting for intra-cavity, fiber coupling, and optic losses, as well as quantum efficiency). We controlled the trap depth via modulation by varying the radio-frequency (RF) power injected by a Quantum Machines OPX+ FPGA into an acousto-optic deflector.

\subsection*{Temperature and Energy Measurements}
We infer the atom’s motional energy along the tweezer axis (\(z\)) by comparing measured cavity transmission to a numerical calibration that maps axial energy to the expected time-averaged transmission. The calibration uses the measured cavity and tweezer parameters to compute the transmitted photon rate for an atom undergoing axial motion \(z(t)\) at fixed energy \(E_z\) in the tweezer potential; we then average the transmission over one oscillation period (and over the photon-counting integration window) to obtain an expected mean transmission fraction. Unless stated otherwise, we neglect radial motion in this calibration since the cavity coupling depends only weakly on the radial degrees of freedom over the populated radial distribution.

These transmissions are compared to the expected time-averaged transmission for an atom oscillating with a fixed longitudinal energy \(E_z\) after assuming that the radial motion is negligible. Specifically, the time-dependent energy shown in Fig.~\ref{fig:fig2}(c) is obtained from a precomputed calibration \(E_z(N,\bar n/\bar n_0)\), where \(\bar n/\bar n_0\) is the normalized transmission and \(N\) is the mean detected photon count in the same time bin. We generate this lookup table by simulating the time-averaged cavity transmission for fixed axial energies over a grid of probe count levels using the measured cavity parameters and the measured empty-cavity level detected counts per bin. For the plotted experimental trace, we first average the surviving trajectories to obtain \(\bar n(t)\), and then evaluate \(E_z\) at each time step. The quantity shown in Fig.~\ref{fig:fig2}(c) should therefore be interpreted as an effective axial-energy estimate obtained from this calibration.

\begin{figure*}
    \centering
    \includegraphics[width=1.0\textwidth]{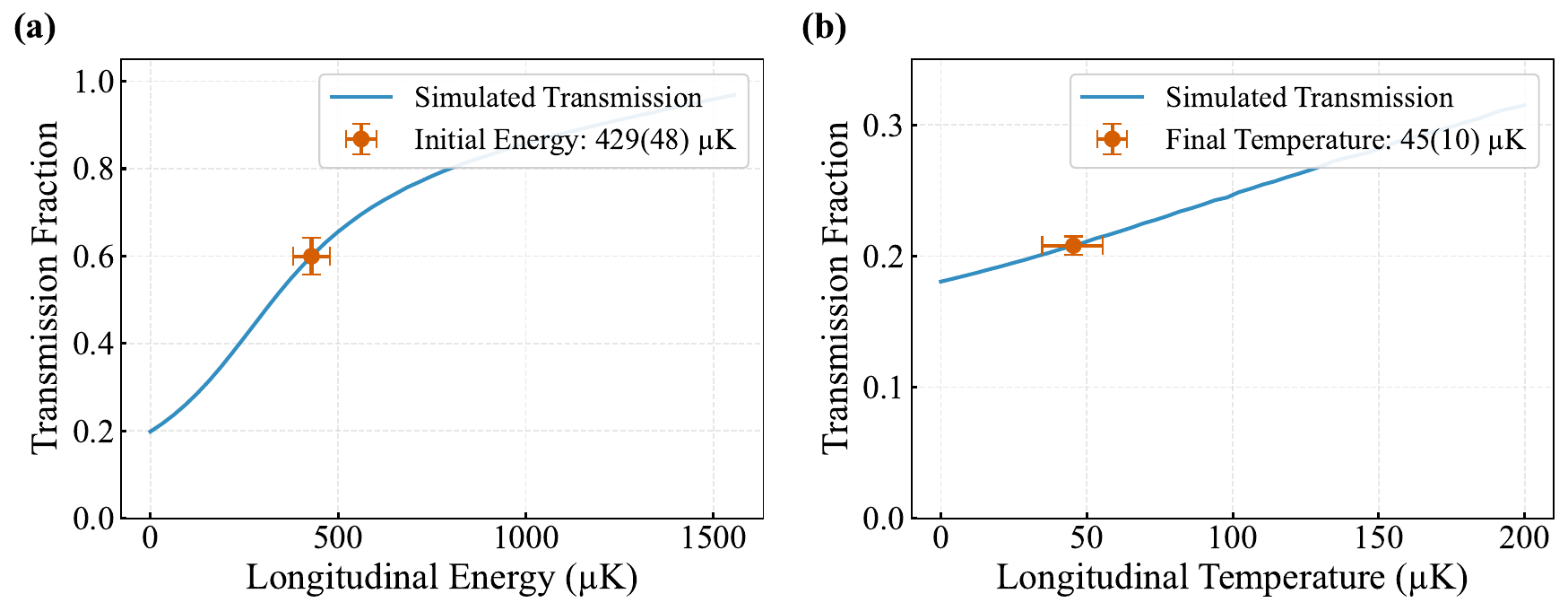}
    \caption{\textbf{Longitudinal Energy and Temperature Calibration.} \textbf{a}, Initial longitudinal energy estimation. The solid blue line shows the simulated transmission fraction as a function of a fixed atomic energy. The experimental data point (orange) represents the measured transmission before feedback is applied, from which we infer an initial energy of \SI{429 \pm 48}{\micro K}. \textbf{b}, Final longitudinal temperature estimation. The solid blue line shows the simulated transmission assuming the atomic longitudinal energy distribution follows a Boltzmann distribution. The experimental data point (orange) corresponds to the transmission measured after feedback cooling, yielding a final temperature of \SI{45 \pm 10}{\micro K}. In both panels, error bars on the data points represent the $1\sigma$ statistical uncertainty.}
    \label{fig:sup_transmission}
\end{figure*}

For the initial calibration point in Fig.~\ref{fig:sup_transmission}(a), the measured transmission is \SI{0.60\pm0.04}{}, from which we infer \(E_z=\SI{429\pm48}{\micro K}\). We first average the transmission and then convert that average into an energy since transmission can be skewed by shot noise to very large values, which would otherwise bias the inferred energy upward. To estimate the post-cooling temperature, we compare the measured transmission averaged over a fixed window after the RL cooling sequence to a second calibration curve computed for a thermal axial energy distribution \(p(E_z)\propto e^{-E_z/(k_B T_z)}\) (Fig.~\ref{fig:sup_transmission}(b)). Using the measured post-cooling transmission \SI{0.208\pm0.007}{}, this yields a final effective axial temperature \(T_z=\SI{45\pm10}{\micro K}\). Error bars denote \(1\sigma\) statistical uncertainties from the finite ensemble size (standard error of the mean propagated through the calibration slope).

For ensemble averages and spectral analyses, we reject trajectories whose smoothed late-time transmission exceeds \(0.85\,\bar n_0\), since this indicates that the atom has been lost and the cavity transmission has returned toward the empty-cavity limit.

\subsection*{Reinforcement Learning Environment Definitions}
\label{appendix:RL_definitions}

\begin{table}[h]
\centering
\begin{tabular}{@{}lll@{}}
\toprule
Symbol & Description & Value \\
\midrule
$N_{frames}$ & Number of frames in state stack & 7 \\
$\beta_f$ & Fast running mean factor & 0.935 \\
$\beta_s$ & Slow running mean factor &  0.99 \\
$w_E$ & Energy reward weight & 7.0 \\
$w_d$ & Difference reward weight & 7.5 \\
$w_b$ & Time bonus reward weight & 0.5 \\
$\alpha_E$ & Reward exponent for $w_E$ & 8 \\
$\Delta t$ & RL agent time step & \SI{27.3}{\micro s} \\
\bottomrule

\end{tabular}
\caption{\textbf{Key environment parameters.} Parameters used to define the reward and observation provided to the RL agent.
}

\label{tab:sim_params}
\end{table}

Our reinforcement learning (RL) environment is a numerical simulator of the coupled atom--cavity dynamics and the photon-count measurement process. Control is discretized into time steps of duration \(\Delta t\); at each step the agent receives an observation derived from recent transmission counts, selects a continuous action that sets the tweezer depth for the next step, and receives a scalar reward. Because the underlying atomic phase-space coordinates are not directly observed, the problem is partially observable. We therefore provide the agent with a short history of engineered features computed from the transmission measurements, which yields near-Markovian behavior for learning.

We employ the Soft Actor-Critic (SAC) algorithm~\cite{SAC_Haarnoja_2018}, a state-of-the-art off-policy method from the actor-critic family known for its high sample efficiency and stability~\cite{AC_Barto_1983, AC_Witten_1997, SAC_Haarnoja_2018}. 
The learning environment is defined by a set of core components. Since the atom's exact phase-space coordinates are unavailable from direct observation due to shot noise, we engineer an observable \textbf{state ($s_t$)} from the history of transmitted photon counts. This state vector provides the agent with proxies for the atom's instantaneous longitudinal position and velocity, derived from the transmitted photon signal, as well as a memory of its most recent action. At each discrete time step, the agent selects an \textbf{action ($a_t$)} to modulate the optical tweezer depth up to its maximum value $U_{\textrm{max}}$, which modifies both the radial and longitudinal optical confining force acting on the atom. The efficacy of this action is then quantified by a scalar \textbf{reward ($r_t$)}, which is engineered to incentivize cooling by rewarding states of low and stable cavity transmission (indicative of low longitudinal atomic motional energy). The agent's objective is finally to learn an optimal \textbf{policy} $\pi_\theta(a_t|s_t)$, a function parametrized by variables $\theta$ that maps states to a distribution over actions. During training, the agent gathers tuples of state, action, and reward into a replay buffer $\mathcal{D}$, from which it repeatedly samples to refine its policy via gradient ascent.

\subsubsection*{Observations, state stack, and action mapping}
At each time step $t$, the SPCM yields an integer photon count $n_t$ integrated over $\Delta t$. We normalize by the mean empty-cavity count $\bar n_0$ (measured with no atom under identical probe conditions) to define a dimensionless transmission signal $c_t$,
\begin{equation}
c_t \equiv \frac{n_t}{\bar n_0}.
\label{eq:ct_def}
\end{equation}
To obtain features with a consistent, centered range for learning, we apply a rescaling factor using calibrated bounds corresponding to an ``empty-cavity'' level and a minimum expected transmission,
\begin{equation}
\tilde c_t \equiv \frac{c_t-c_{\min}}{c_{\max}-c_{\min}},
\label{eq:ct_affine}
\end{equation}
with $c_{\max}\!=\!1$ and $c_{\min}$ set by the calculated expected minimum transmission under maximal atom--cavity coupling. 

We compute exponential moving averages of $\tilde c_t$ on two timescales to form fast and slow baselines, with coefficients $\beta_f$ and $\beta_s$ chosen such that the fast baseline responds more rapidly than the slow baseline,
\begin{align}
\tilde m_t^{(f)} &= \beta_f\, \tilde m_{t-1}^{(f)} + (1-\beta_f)\,\tilde c_t, \\
\tilde m_t^{(s)} &= \beta_s\, \tilde m_{t-1}^{(s)} + (1-\beta_s)\,\tilde c_t.
\label{eq:ema_defs}
\end{align}
We additionally define a difference trend feature
\begin{equation}
\tilde d_t \equiv \tilde m_t^{(s)}-\tilde m_t^{(f)},
\label{eq:dt_def}
\end{equation}
which is positive when the transmission is decreasing relative to the slow baseline (a signature of cooling) and negative when the transmission is rising (heating).

Next, the continuous action $a_t\in[-1,1]$ selected by the agent is mapped to a physical tweezer depth $U_t\in[0,U_{\mathrm{max}}]$ according to a quadratic transfer function that accounts for the measured nonlinear AOD response and yields an approximately linear relation between $a_t$ and delivered optical power,
\begin{equation}
U_t = U_{\mathrm{max}}\left(\frac{a_t+1}{2}\right)^2.
\label{eq:action_map}
\end{equation}

Based on these features, we assemble the per-time-step observation vector
\begin{equation}
o_t \equiv \big(a_{t-1},\; \tilde c_t-\tfrac{1}{2},\; \tilde m_t^{(f)}-\tfrac{1}{2},\; 2\tilde d_t\big),
\label{eq:obs_def}
\end{equation}
where the centering by $1/2$ and the factor of $2$ on the trend term are chosen to keep all components roughly in the interval $[-1,+1]$. To ensure the inputs to the agent can be approximated as Markovian, the input state at time $t$ is defined as a stack of the most recent $N_{\mathrm{frames}}$ observation vectors,
\begin{equation}
s_t \equiv (o_{t-N_{\mathrm{frames}}+1},\,o_{t-N_{\mathrm{frames}}+2},\,\ldots,\,o_t),
\label{eq:state_stack}
\end{equation}
which in effect provides the agent a memory.

\subsubsection*{Reward function}
The reward is designed to encourage rapid reduction of the atom's longitudinal motional energy while maintaining confinement. Because the transmission measurement is shot-noise limited, we compute rewards from the filtered transmission features defined above rather than directly from the raw count $n_t$.

The per-step reward returned after applying action $a_t$ is
\begin{equation}
\begin{split}
r_{t+1} = & \ w_E\,\mathbf{1}_{t\ge t_{\mathrm{warm}}}\big(\mathrm{clip}(\tilde m_t^{(f)},0,1)-1\big)^{\alpha_E} \\
& + w_d\,\tilde d_t + w_b\,\mathbb{I}[\mathrm{trapped}]
\end{split}
\label{eq:reward_def}
\end{equation}
where $\mathbb{I}[\mathrm{trapped}]$ is 1 if the atom remains trapped during the step and 0 otherwise, and $\mathrm{clip}(\cdot,0,1)$ truncates its argument to $[0,1]$. The exponent $\alpha_E$ is chosen even so that the low-transmission term is non-negative. The warm-up index $t_{\mathrm{warm}}$ delays the application of the nonlinear low-transmission term until the fast exponential moving average has reached steady-state; in implementation we set
\begin{equation}
t_{\mathrm{warm}}=\left\lfloor\frac{1}{1-\beta_f}\right\rfloor,
\end{equation}
corresponding to approximately one effective time constant of the fast filter.

In the simulator used for this work, the weights $(w_E,w_d,w_b)$ are tuned empirically for stable learning. The first term provides a strong incentive to reach and remain at low mean transmission (tight localization), the trend term $w_d\,\tilde d_t$ rewards decreasing transmission (cooling) and penalizes increasing transmission (heating), and the survival bonus $w_b$ favors policies that maintain trapping over longer horizons. Episodes terminate early upon atom loss and otherwise at a fixed horizon $t_{\max}$, which removes future reward and enforces the survival--cooling trade-off. Key parameters for the reward and state definitions in the environment are shown in Table \ref{tab:sim_params}.

\subsubsection*{System dynamics}
The atom's trajectory is modeled by a set of stochastic differential equations (SDEs) describing the evolution of its phase-space coordinates $Y(t) \equiv (\vec{r},\vec{v},a_r,a_i)$, where position is given by $\vec{r}=(x,y,z)$, velocity by $\vec{v}=(v_x,v_y,v_z)$, and the real and imaginary parts of the intracavity field amplitude given by $\alpha = a_r + i a_i$~\cite{EoMSim_Ritsch_2001}. We treat the atom as a two-level system that dispersively couples to the cavity field. The general form of the SDE used in the simulation is
\begin{equation}
\dd{Y} = F(Y,t)\,\dd{t} + G(Y,t)\dd{W}_t,
\label{eq:sde_compact}
\end{equation}
where \(F\) is the deterministic drift, \(G\) is a state-dependent diffusion matrix, and \(\dd{W}_t\) is a vector of independent Wiener increments. Expanding the drift terms gives
\begin{align}
\dot{\vec r} &= \vec v, \label{eq:eom_r}\\
m\,\dot{\vec v} &= \vec F_{\mathrm{trap}}(\vec r,t) + \vec F_{\mathrm{cav}}(\vec r,\alpha) + \vec F_{\mathrm{sc}}(\vec r,\alpha), \label{eq:eom_v}\\
\dot a_r &= -E + \big(U_{0,a}(\vec r,\alpha)-\delta\big)a_i - \Big(\frac{\kappa}{2}+\Gamma_a(\vec r,\alpha)\Big)a_r, \label{eq:eom_ar}\\
\dot a_i &= -\big(U_{0,a}(\vec r,\alpha)-\delta\big)a_r - \Big(\frac{\kappa}{2}+\Gamma_a(\vec r,\alpha)\Big)a_i. \label{eq:eom_ai}
\end{align}
Here \(m\) is the atomic mass and \(E\) is the (real) probe drive amplitude. The terms \(U_{0,a}(\vec r,\alpha)\) and \(\Gamma_a(\vec r,\alpha)\) denote, respectively, the atom-induced dispersive cavity shift and the atom-induced loss rate (free-space scattering) in our semiclassical model.

The atomic force in Eq.~\eqref{eq:eom_v} is decomposed into: (i) the conservative optical-tweezer force \(\vec F_{\mathrm{trap}}=-\nabla U_{\mathrm{trap}}(\vec r,t)\), where the trap depth is set by the RL action via Eq.~\eqref{eq:action_map}, (ii) the conservative cavity dipole force \(\vec F_{\mathrm{cav}}\), and (iii) the dissipative scattering (radiation-pressure) force \(\vec F_{\mathrm{sc}}\). The diffusion matrix \(G(Y,t)\) injects fluctuations into the atomic velocities and the field quadratures to model momentum diffusion and cavity shot noise.

\subsection*{Experimental Training and Architecture}

\begin{figure}
    \centering
    \includegraphics[width=0.5\textwidth]{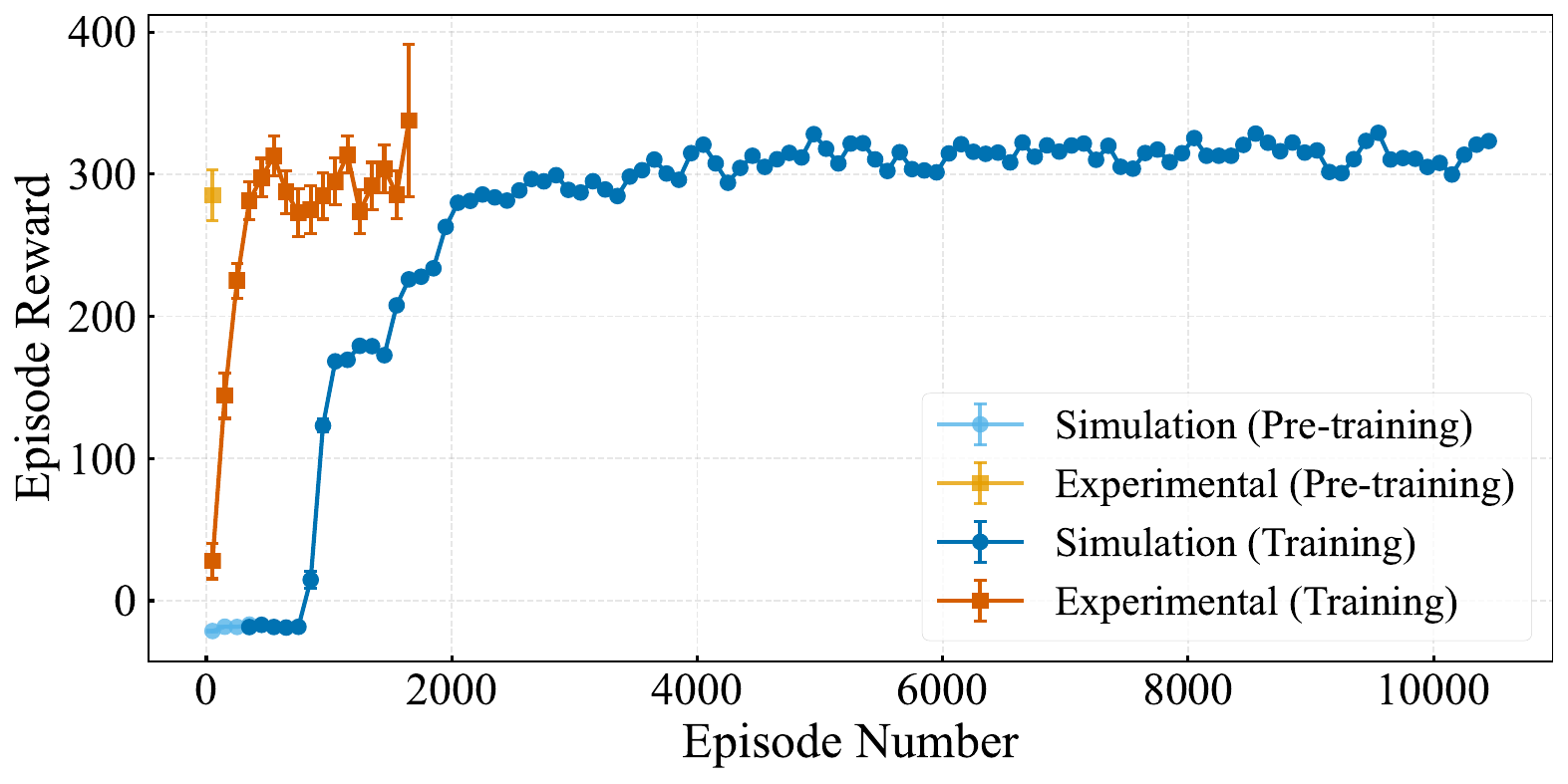}

        \caption{\textbf{ \emph{Sim-to-Real} Training }  Episode-reward learning curves for policies evaluated purely in simulation (blue) and on the experimental apparatus (orange). In both cases, there is an initial pre-training stage where an agent explores the environment with either a randomly initialized policy in simulation or the replay-buffer-seeded policy in the experiment. 
        Light-shaded markers show the performance of the policies during pre-training, while dark-shaded markers indicate the subsequent on-line training phase. Pre-training encompasses collecting data without changes to the policy. The experimental agent (orange) surpasses 90\% of its final reward after only $\sim$350 experimental episodes, whereas learning from scratch in simulation requires $\gtrsim2250$. }

\label{fig:supp_training}
\end{figure}

\subsection*{Neural Network Structure}
The agent employs an actor-critic architecture, where the actor and critic are separately parameterized by neural networks. This structure allows the agent to simultaneously learn an optimal control strategy and estimate the long-term value of its actions.
\begin{itemize}
    \item \textbf{Actor Policy Network:} The policy network, which learns the control mapping $\pi_\theta(a_t|s_t)$, is a multi-layer perceptron (MLP) with two fully connected hidden layers of 16 neurons, set by hardware constraints. It outputs the mean and standard deviation of a Gaussian distribution from which a continuous action is sampled.
    \item \textbf{Critic Network:} The action-value network, which learns the Q-function, is a MLP with three fully connected hidden layers of 512, 256, and 128 neurons, respectively. 
\end{itemize}
For both networks, all hidden layers utilize the Leaky Rectified Linear Unit (LeakyReLU) as their activation function. The network weights are optimized using the AdamW algorithm \cite{AdamW_Hutter_2019}.

\subsubsection*{Simulated Training and Evaluation Protocol}

Agents trained on the simulation learn for a total of $N_{\textrm{total}} = 1,020,000$ time steps, and a single episode consists of at most 100 time steps before the environment is reset. 100 time steps were chosen as this corresponds to several multiples of the average cooling time, which allows the system to settle into a steady state. A sample training curve is shown in Fig.~\ref{fig:supp_training} for simulation (darker colors). As an off-policy algorithm, SAC learns from a large replay buffer, $\mathcal{D}$, which stores the $N_{\mathcal{D}} = 500,000$ most recent interaction tuples. The learning process begins only after an initial $N_{\textrm{warm-up}} = 10,000$ ``warm-up'' steps have been collected to ensure the buffer is adequately populated with diverse experiences. After every $T_{\textrm{update}} = 256$ steps of environmental interaction, the agent performs $N_{\textrm{grad}} = 256$ gradient-descent updates with learning rate $l_r=10^{-4}$ for both actor and critic, with each update using a mini-batch of size $N_{\textrm{batch}} = 256$ randomly sampled from the replay buffer. The target Q-network is updated after $T_{\textrm{target}}$ environmental steps.

A central feature of the SAC algorithm is its use of an entropy maximization framework to balance the exploration-exploitation dilemma. The agent's objective function includes a term that rewards higher policy entropy, encouraging the agent to explore a wider range of actions and preventing premature convergence to a suboptimal policy. The relative importance of this entropy term is controlled by a temperature parameter, the entropy coefficient $\chi$, which is automatically tuned during training to maintain an optimal balance~\cite{stable-baselines3_Dormann_2021}. Key hyperparameters for the training protocol are summarized in Table~\ref{tab:rl_params}, and their values were determined by a random hyperparameter search over many different runs.

To estimate a systematic uncertainty lower-bound in the light-blue shaded region in Fig. \ref{fig:fig3} associated with experimental calibration and parameter drift, we repeated the simulation sweep with randomized noisy experimental parameters. For each trajectory, the photon-count calibration was sampled with a Gaussian fractional uncertainty of $25\%$, and the tweezer depth was sampled with a Gaussian fractional uncertainty of $10\%$. We also included transverse trap-cavity misalignment by drawing independent offsets in the two transverse directions from Gaussian distributions with rms width \SI{2.03}{\micro m}, corresponding approximately to a $15\%$ rms variation in cooperativity. The radial temperature was set to \SI{180}{\micro K} to account for sensitivity to alignment and imperfect knowledge of the transverse motional distribution. The simulated photon detection efficiency in this noisy run was set to half the detector efficiency, and photon counts were sampled from a Poisson distribution. Cavity detuning noise was included as a correlated step-to-step drift with scale $\kappa/6$.

\subsection*{Experimental Training Procedure}

\begin{table}[h]
\centering
\begin{tabular}{@{}lll@{}}
\toprule
\textbf{Hyperparameter} & \textbf{Symbol} & \textbf{Value} \\
\midrule
\multicolumn{3}{@{}l}{\textbf{SAC Algorithm Parameters}} \\
Optimizer & - & AdamW \\
Learning Rate & $l_r$ & \num{1e-4} \\
Discount Factor & $\gamma$ & 0.98 \\
Polyak-averaging Factor & $\tau$ & 0.01 \\
Replay Buffer Size & $N_{\mathcal{D}}$ & 500,000 \\
Mini-batch Size & $N_{\textrm{batch}}$ & 256 \\
Target Update Interval & $T_{\textrm{target}}$ & 4 \\
Entropy Coefficient & $\chi$ & Auto-tuned (see \cite{stable-baselines3_Dormann_2021}) \\
\midrule
\multicolumn{3}{@{}l}{\textbf{Simulation Training and Evaluation Parameters}} \\
Total Timesteps & $N_{\textrm{total}}$ & 1,020,000 \\
Learning Starts & $N_{\textrm{warm-up}}$ & 10,000 \\
Training Update Frequency & $T_{\textrm{update}}$ & 256 steps \\
Gradient Steps per Update & $N_{\textrm{grad}}$ & 256 \\

\bottomrule
\end{tabular}
\caption{\textbf{Key RL training hyperparameters.} A summary of the main hyperparameters used for training the Soft Actor-Critic (SAC) agent in the simulated environment.}
\label{tab:rl_params}
\end{table}

To transfer the knowledge from simulation to the physical system, we initialize experimental training by seeding the agent's replay buffer $\mathcal{D}$ with approximately $6000$ timesteps of data collected by this pre-trained policy on the experiment (lighter colors; Fig.~\ref{fig:supp_training}). The experimental agent is then initialized with a random policy for experimental training (dark orange), and learns from both new data collected and the simulation-trained agent's pre-seeded data. This seeding dramatically accelerates the learning process: the agent reaches 90\% of its final reward in just $\sim 350$ episodes, a substantial improvement over the $\sim 2250$ episodes required when training from scratch in simulation. The simulation-trained policy already achieves $\sim 93\%$ of the final, optimized cumulative experimental reward, which indicates the high fidelity of our simulation.

The online training protocol for the physical experiment incorporates several modifications to adapt to real-world constraints over the simulation. Firstly, it is not possible to update experimental network parameters within an episode as it is for the simulation. The experimental logic is therefore modified to trigger a network update only after \textit{at least} $T_{\textrm{update}}$ new timesteps have been accumulated across one or more episodes. To prevent the policy from remaining stuck where it causes immediate atom loss (and no updates occur), we additionally implement a training step after 12 consecutive experimental runs with no updates. Furthermore, the simulation explicitly models the $\sim\SI{7.3}{\micro s}$ computational latency of our FPGA controller, which trains the simulated agent to be robust to the delay between observation and action in the physical system. This latency delays the action between observation of the system's state and modulation of the dipole trap power, and additionally creates SPCM dead time during which photon counts are not recorded. Finally, the initialization of the agent's observational history differs slightly. In simulation, the agent begins with a pre-populated history corresponding to constant trap power, whereas in the experiment, the agent must build its observational history from scratch at the beginning of each episode. The training procedure between simulation and experiment is otherwise identical.

\section*{Data availability}
The full datasets generated and analysed during the current study are available from the corresponding author upon reasonable request.

\section*{Code availability}
The code and source data needed to reproduce the results are available at \url{https://github.com/math2peters/deep_RL_cavity_feedback_cooling}.

\bibliography{main}

\section*{Author Contributions}
M. Peters, G. Wang, and D. Spierings contributed equally to this work. M. Peters, V. Vuletić, and I. Chuang conceived the experiment. M. Peters, G. Wang, and D. Spierings designed and built the experimental apparatus. M. Peters and N. Drucker developed the real-time feedback-control implementation and FPGA/experimental control interface. M. Peters developed the atom–cavity simulation environment. M. Peters developed and trained the reinforcement-learning agents. M. Peters and G. Wang performed the experimental measurements. M. Peters, M. Chen, and A. Bartlett analysed the data. M. Peters and I. Chuang developed the theoretical model and physical interpretation of the learned control policy. M. Peters wrote the manuscript with input from all authors. All authors discussed the results and contributed to the final manuscript.

\section*{Acknowledgements}
This work is supported by the U.S. Department of Energy, Office of Science, National Quantum Information Science Research Centers, Quantum Systems Accelerator. Additional support is acknowledged from the NSF Frontier Center for Ultracold Atoms (grant number PHY-2317134), the NSF QLCI Q-SEnSE (grant number QLCI-2016244), and the NSF National Quantum Virtual Laboratory (NQVL) (grant number OSI-2533041).
We thank Berk Kovos and Christian Hahn for their helpful discussions on integrating the Quantum Machines OPX+ device into our setup. We thank Uroš Delić for valuable discussions in cavity cooling and model-based cooling. We gratefully acknowledge the MIT SuperCloud team and the Stable Baselines3 implementation of RL agent training software~\cite{stable-baselines3_Dormann_2021}.

\section*{Competing Interests}
Quantum Machines provided the OPX+ hardware used for real-time feedback control in this work. N. Drucker is an employee of Quantum Machines. The remaining authors declare no competing interests.

\clearpage
\onecolumngrid

\appendix
\section*{Supplementary Information}
\setcounter{equation}{0}
\setcounter{figure}{0}
\setcounter{table}{0}
\renewcommand{\theequation}{S\arabic{equation}}
\renewcommand{\thefigure}{S\arabic{figure}}
\renewcommand{\thetable}{S\arabic{table}}

\label{supplemental_materials}

\section*{Passive Cooling Mechanism}\label{secA1}
In the following sections, we derive equations for the passive cooling mechanism the atom experiences within the cavity mode due to interactions with intracavity photons. We first treat the case where the atom is aligned to the center of the cavity mode followed by the case where the atom is offset from the cavity mode.

In the dispersive and low saturation regime, the equation of motion for the cavity field can be written as

\begin{equation}
\begin{aligned}
\frac{da}{dt}
&=\left[i(\delta-U(t))-\frac{\kappa}{2}\left(1+\eta(t)\mathcal{L}_a(\Delta)\right)\right]a-\epsilon \\
&\approx \left[i(\delta-U(t))-\frac{\kappa}{2}\right]a-\epsilon ,
\end{aligned}
\end{equation}
where $\epsilon$ is the (real) drive strength, $\eta(t)=\frac{4g(t)^2}{\kappa\Gamma}$ is the time-dependent cooperativity, $\mathcal{L}_a(\Delta)=\frac{\Gamma^2}{\Gamma^2+4\Delta^2}$ is the absorptive lineshape, and $U(t) =\frac{g(t)^2\Delta}{(\frac{\Gamma}{2})^2+\Delta^2}\approx \frac{g(t)^2}{\Delta}$ is the time-dependent dispersive shift due to the atomic motion. Along the axial, $z$-direction of the tweezer, the potential can be written as $U_0e^{-\frac{2z(t)^2}{w_c^2}}$ where $U_0=g_{\mathrm{max}}^2/\Delta$ is the dispersive shift at the center of the cavity and $z(t)$ is the $z$-position of the atom. 

\subsection*{Tweezer center aligned to cavity center}
Let us assume that a single atom $N=1$ is trapped in an optical tweezer with a depth (on the order of 20~MHz), much larger than the dispersive shift (on the order of 10~kHz), and consider the axial motion of the atom in the tweezer with a trapping frequency of $\omega_t$. We assume the motion of the atom approximately follows harmonic oscillation with a slowly decaying amplitude $A(t)$. The atomic position along the axial direction is then $z(t)=A(t)\sin(\omega_t t)$, where $A(t)$ can be treated quasi-statically $A(t)\approx A$ during an oscillation period. When the oscillation amplitude is smaller than the cavity waist, we can expand the form of $U(t)$ and approximately get $U(t)\approx U_0[(1-\frac{A^2}{w_c^2})+\frac{A^2}{w_c^2}\cos(2\omega_t t)]$ which contains a DC component and an AC component with an oscillation amplitude of $U_0A^2/w_c^2$.
The equation of motion is then written as

\begin{equation}
\frac{da}{dt}
=
\left[
i\!\left(\delta-U_0\!\left[\left(1-\frac{A^2}{w_c^2}\right)+\frac{A^2}{w_c^2}\cos(2\omega_t t)\right]\right)
-\frac{\kappa}{2}
\right]a-\epsilon .
\label{eq:dadt}
\end{equation}
Assuming a quasi-steady-state solution of the form
\begin{equation}
a(t) = a_0 + a_2 e^{i 2\omega_t t} + a_{-2} e^{-i 2\omega_t t},
\end{equation}
we can plug into Eq.~\eqref{eq:dadt} to solve for the coefficients $a_0, a_{\pm 2}$ and get a steady-state solution with oscillations at $2\omega_t$

\begin{equation}
\begin{aligned}
a(t) &= \frac{\epsilon}{i (\delta - U_0 (1 - \frac{A^2}{w_c^2})) - \frac{\kappa}{2}} \\
&\quad + \frac{i U_0 \frac{A^2}{2w_c^2} a_0}{i (\delta-2\omega_t - U_0 (1 - \frac{A^2}{w_c^2})) - \frac{\kappa}{2}} e^{i 2\omega_t t} \\
&\quad + \frac{i U_0 \frac{A^2}{2w_c^2} a_0}{i (\delta+2\omega_t - U_0 (1 - \frac{A^2}{w_c^2})) - \frac{\kappa}{2}} e^{-i 2\omega_t t} .
\end{aligned}
\end{equation}
The intracavity photon number 

\begin{equation}
\begin{aligned}
n(t) &\equiv \langle a^\dagger(t)a(t)\rangle \\
&\approx \frac{\epsilon^2}{(\frac{\kappa}{2})^2+\tilde{\delta}^2} \\
&\quad + 2[\Re(a_0^*a_2)+\Re(a_0 a_{-2}^*)]\cos(2\omega_t t) \\
&\quad - 2[\Im(a_0^*a_2)+\Im(a_0a_{-2}^*)]\sin(2\omega_t t) .
\end{aligned}
\end{equation}
Here we define a detuning relative to the dispersively shifted cavity resonance $\tilde{\delta}=\delta-U_0(1-\frac{A^2}{w_c^2})$. The $\cos(2\omega_t t)$ term follows the steady-state response of the cavity intensity. In contrast, the $\sin(2\omega_t t)$ term has a 90-degree phase shift, thus can lead to energy dissipation which we will discuss shortly. We can further express

\begin{align}
    \Im(a_0^*a_2)&=-|a_0|^2\frac{A^2}{2w_c^2}\frac{U_0\frac{\kappa}{2}}{(\tilde{\delta}-2\omega_t)^2+\frac{\kappa^2}{4}},\\
    \Im(a_0a_{-2}^*)&=|a_0|^2\frac{A^2}{2w_c^2}\frac{U_0\frac{\kappa}{2}}{(\tilde{\delta}+2\omega_t)^2+\frac{\kappa^2}{4}},\\
    \Im(a_0^*a_2)+\Im(a_0a_{-2}^*)&=-|a_0|^2\frac{A^2}{w_c^2}F(\omega_t,\delta)\text{, with }\\
    F(\omega_t,\delta)&=\frac{2\Tilde{\delta}\omega_t\kappa U_0}{(\tilde{\delta}^2+4\omega_t^2+\frac{\kappa^2}{4})^2-16\tilde{\delta}^2\omega_t^2}.
    \label{eq:imaginary}
\end{align}
When setting $\delta=0$ such that $\tilde{\delta}=-U_0(1-\frac{A^2}{w_c^2})$, we see that the fractional function due to phase delay $F(\omega_t,0)<0$ is negative. The dipole force that the atom experiences from intracavity photons is then given by

\begin{equation}
\begin{aligned}
f(t) &= -\frac{\partial}{\partial z}[n(t)U(t)] \\
&= \frac{4}{w_c^2}z(t)n(t)U(t) \\
&= \frac{4AU_0}{w_c^2}\sin(\omega_t t)\left[\left(1-\frac{A^2}{w_c^2}\right)+\frac{A^2}{w_c^2}\cos(2\omega_t t)\right]n(t) .
\end{aligned}
\end{equation}
We note that here we did not consider the dipole force generated by the tweezer itself, under the assumption that the potential of the tweezer is conservative and does not lead to energy loss.
The velocity of the atom is 
\begin{equation}
    v(t)=A\omega_t\cos(\omega_t t).
\end{equation}
The energy change of the atom over a period of oscillation can be calculated by integrating

\begin{align}
\Delta E
&= \int_0^T dt\, f(t)\,v(t) \\
&= \frac{4A^2\omega_t U_0}{w_c^2}
   \int_0^T dt\,
   \sin(\omega_t t)\cos(\omega_t t)\notag\\
&\qquad\times
   \left[
      \left(1-\frac{A^2}{w_c^2}\right)
      + \frac{A^2}{w_c^2}\cos(2\omega_t t)
   \right] n(t) \\
&= \frac{2A^2\omega_t U_0}{w_c^2}\times \int_0^T dt\ 
   \left[
      \left(1-\frac{A^2}{w_c^2}\right)\sin(2\omega_t t)
      + \frac{A^2}{2w_c^2}\sin(4\omega_t t)
   \right] n(t) \\
&= -2\pi \frac{A^2}{w_c^2} U_0
   \left(1-\frac{A^2}{w_c^2}\right)\notag \times
   2\!\left[
      \Im(a_0^* a_2) + \Im(a_0 a_{-2}^*)
   \right] \\
&= 4\pi U_0
   \left(\frac{A^2}{w_c^2}\right)^2
   \left(1-\frac{A^2}{w_c^2}\right)
   |a_0|^2\,F(\omega_t,\delta).
\end{align}

The energy change per period is contributed from the maximum kinetic energy that the atom can dissipate $U_0\frac{A^2}{w_c^2}(1-\frac{A^2}{w_c^2})$ assuming ``perfect" phase delay for oscillation damping for each single photon in the cavity, the maximum fractional photon number change in the cavity leading to the intensity fluctuation $\frac{A^2}{w_c^2}|a_0|^2$, as well as a dimensionless function characterizing the phase-delay effect $F(\omega_t,\delta)$. When the probe is on empty-cavity resonance, we have $F<0$ leading to cavity cooling. Changing the detuning $\delta$ thus changes the effect from cavity cooling to heating.

Further analyzing the time-dependence of the atomic energy $E=\frac{mA^2\omega_t^2}{2}$, we have 
\begin{equation}
\frac{dE}{dt}\approx 2\omega_t U_0 \frac{E^2}{w_c^4} (\frac{2}{m\omega_t^2})^2 |a_0|^2\times F(\omega_t, \delta)
\end{equation}
yielding
\begin{equation}
    E(t)=E_0\frac{1}{1-2E_0\omega_t U_0 (\frac{2}{m\omega_t^2w_c^2})^2|a_0|^2 F(\omega_t,\delta) t},
\end{equation}
here we assume $(1-\frac{A^2}{w_c^2})\approx 1$ so it neatly describes the cooling behavior in the final stage. 
Using the fact that axial trapping frequency of a tweezer is $\omega_t=\sqrt{\frac{2U_t}{mz_R^2}}$ where $U_t$ is the trap depth of the tweezer, the decay of the energy can be further written as
\begin{equation}
    E(t)=E_0\frac{1}{1-\omega_t \frac{z_R^4}{w_c^4}\frac{E_0U_0}{4U_t^2}|a_0|^2 F(\omega_t,\delta) t}.
\end{equation}

\subsection*{Tweezer center is not aligned to the cavity center}
When the tweezer center is offset from the cavity center along the axial direction, the potential can be written as $U_0e^{-\frac{2(z(t)-z_0)^2}{w_c^2}}=U_0e^{-\frac{2z(t)^2}{w_c^2}+\frac{4z(t)z_0}{w_c^2}-\frac{2z_0^2}{w_c^2}}=(U_0e^{-\frac{2z_0^2}{w_c^2}})e^{-\frac{2z(t)^2}{w_c^2}+\frac{4z(t)z_0}{w_c^2}}$, we can still derive the cavity field with
\begin{align}
    \frac{da}{dt}&=\left[i(\delta-U_0^\prime e^{-\frac{2z(t)^2}{w_c^2}+\frac{4z(t)z_0}{w_c^2}})-\frac{\kappa}{2}\right]a-\epsilon\\
    &\approx\left[i\left(\delta-U_0^\prime\left(1-\frac{A^2}{w_c^2}+\frac{A^2}{w_c^2}\cos(2\omega_t t)+\frac{4z_0 A}{w_c^2}\sin(\omega_t t)\right)\right)-\frac{\kappa}{2}\right]a-\epsilon.
\end{align}
Here we assume that the offset of the tweezer center is small compared to the cavity waist, while we note that a larger offset could be treated as well by keeping more terms in the Taylor expansion of the exponential function. 

The major difference from the previous case is that the quasi-steady state of the cavity field has additional first-order frequency terms $\omega_t$, which can be described by 
\begin{equation}
a(t) = a_0 + a_{+1} e^{i\omega_t t} + a_{-1} e^{-i\omega_t t} + a_{+2} e^{i2\omega_t t} + a_{-2} e^{-i2\omega_t t}.
\end{equation}
The intracavity photon number can then be written as
\begin{align}
n(t) &\approx |a_0|^2 + |a_{+1}|^2 + |a_{-1}|^2 + |a_{+2}|^2 + |a_{-2}|^2 \nonumber \\
&\quad + 2 \left( \Re(a_0 a_{+1}^*) + \Re(a_0 a_{-1}^*) \right) \cos\omega_t t \nonumber \\
&\quad + 2 \left( \Im(a_0 a_{+1}^*) - \Im(a_0 a_{-1}^*) \right) \sin\omega_t t \nonumber \\
&\quad + 2 \left( \Re(a_0 a_{+2}^*) + \Re(a_0 a_{-2}^*)) \right) \cos 2\omega_t t \nonumber \\
&\quad + 2 \left( \Im(a_0 a_{+2}^*) - \Im(a_0 a_{-2}^*)) \right) \sin 2\omega_t t.
\end{align}
In the following, we will show that the dominant term that leads to energy dissipation is the $\cos(\omega_t t)$ term with an amplitude
\begin{equation}
    \Re(a_0a_1^*)+\Re(a_0a_{-1}^*)=|a_0|^2\frac{z_0A}{w_c^2}\frac{16\kappa U_0^\prime (\kappa^2+4\tilde{\delta}^2+4\omega_t^2)}{(\kappa^2+4\tilde{\delta}^2+4\omega_t^2)^2+(8\tilde{\delta}\omega_t)^2}=|a_0|^2\frac{z_0 A}{w_c^2}F^\prime(\omega_t,\delta).
\end{equation}

The dipole force that the atom experiences is given by
\begin{align}
    f(t)=-\frac{\partial}{\partial z} n(t)U(t)&=\frac{4(z(t)-z_0)}{w_c^2} n(t)U(t)\\&=\frac{4AU_0^\prime}{w_c^2}\sin(\omega_t t)\left[(1-\frac{A^2}{w_c^2})+\frac{A^2}{w_c^2}\cos(2\omega_t t)+\frac{4z_0A}{w_c^2}\sin(\omega_t t)\right]n(t)\\&-\frac{4z_0U_0^\prime}{w_c^2}\left[(1-\frac{A^2}{w_c^2})+\frac{A^2}{w_c^2}\cos(2\omega_t t)+\frac{4z_0A}{w_c^2}\sin(\omega_t t)\right]n(t)
\end{align}
The velocity of the atom is still
\begin{equation}
    v(t)=A\omega_t\cos(\omega_t t).
\end{equation}
The energy change of the atom over a period of oscillation can be calculated with
\begin{align}
    \Delta E&=\int_0^T dt f(t)v(t)\\&=\int_0^T dt \omega_t  \bigg[\frac{4A^2U_0^\prime}{w_c^2}\sin(\omega_t t)\cos(\omega_t t)\left[(1-\frac{A^2}{w_c^2})+\frac{A^2}{w_c^2}\cos(2\omega_t t)+\frac{4z_0A}{w_c^2}\sin(\omega_t t)\right]n(t)\\&\ \ -\frac{4z_0AU_0^\prime}{w_c^2}\cos(\omega_t t)\left[(1-\frac{A^2}{w_c^2})+\frac{A^2}{w_c^2}\cos(2\omega_t t)+\frac{4z_0A}{w_c^2}\sin(\omega_t t)\right]n(t)\bigg]\\
    &=\int_0^T dt \omega_t  \bigg[\frac{2A^2U_0^\prime}{w_c^2}\left[(1-\frac{A^2}{w_c^2})\sin(2\omega_t t)+\frac{A^2}{2w_c^2}\sin(4\omega_t t)+\frac{4z_0A}{w_c^2}\sin(2\omega_t t)\sin(\omega_t t)\right]n(t)\\&\ \ -\frac{4z_0AU_0^\prime}{w_c^2}\left[(1-\frac{A^2}{w_c^2})\cos(\omega_t t)+\frac{A^2}{w_c^2}\cos(\omega_t t)\cos(2\omega_t t)+\frac{2z_0A}{w_c^2}\sin(2\omega_t t)\right]n(t)\bigg]\\
    &\approx 4\pi \frac{A^2U_0^\prime}{w_c^2}(1-\frac{A^2}{w_c^2})(\Im(a_0a_{2}^*)-\Im(a_0a_{-2}^*)) +8\pi\frac{A^2U_0^\prime}{w_c^2}\frac{z_0A}{w_c^2}(\Re(a_0a_{+1}^*)+\Re(a_0a_{-1}^*)) \\
    &-8\pi\frac{z_0 AU_0^\prime}{w_c^2}\left[(1-\frac{A^2}{w_c^2})(\Re(a_0a_{+1}^*)+\Re(a_0a_{-1}^*))+\frac{2z_0 A}{w_c^2}(\Im(a_0a_{2}^*)-\Im(a_0a_{-2}^*)) \right].
\end{align}
Keeping terms in the lowest order of $A/w_c$, we can obtain
\begin{equation}
    \Delta E\approx -8\pi\frac{z_0AU_0^\prime}{w_c^2} (\Re(a_0a_1^*)+\Re(a_0a_{-1}^*))=-8\pi \frac{z_0 A }{w_c^2}U_0^\prime\times \frac{z_0 A}{w_c^2}|a_0|^2\times F^\prime(\omega_t,\delta).
\end{equation}
Now instead the energy change depends quadratically on the oscillation amplitude of the atomic position, thus an exponential decay of energy is expected.
Further analyzing the time-dependence of the atomic energy $E=\frac{mA^2\omega_t^2}{2}$, we can obtain the time dependence 
\begin{equation}
\frac{dE}{dt}\approx -4\omega_t U_0^\prime E\frac{z_0^2}{w_c^4} (\frac{2}{m\omega_t^2})|a_0|^2 \times F^\prime(\omega_t, \delta)
\end{equation}
yielding
\begin{equation}
    E(t)=E_0\exp(-4\omega_t |a_0|^2 \frac{z_0^2}{w_c^4} U_0^\prime (\frac{2}{m\omega_t^2})F^\prime(\omega_t,\delta)t).
\end{equation}
We note that the axial trapping frequency of a tweezer is $\omega_t=\sqrt{\frac{2U_t}{mz_R^2}}$, the decay of the energy can be further written as
\begin{equation}
    E(t)=E_0\exp(-4\omega_t |a_0|^2\frac{z_0^2z_R^2}{w_c^4}\frac{U_0^\prime}{U_t}F^\prime(\omega_t,\delta)t).
\end{equation}

\end{document}